\documentclass[preprint]{elsarticle}
\usepackage{amsmath}
\usepackage{amssymb}
\usepackage{color}
\usepackage{graphicx}
\usepackage[]{enumerate}
\usepackage{bm}

\DeclareBoldMathCommand{\bV}{V}
\DeclareBoldMathCommand{\bv}{v}
\DeclareBoldMathCommand{\bx}{x}
\DeclareBoldMathCommand{\by}{y}
\DeclareBoldMathCommand{\bz}{z}
\DeclareBoldMathCommand{\bF}{F}
\DeclareBoldMathCommand{\br}{r}
\DeclareBoldMathCommand{\bb}{b}
\DeclareBoldMathCommand{\be}{e}
\DeclareBoldMathCommand{\bB}{B}
\DeclareBoldMathCommand{\bE}{E}
\DeclareBoldMathCommand{\bk}{k}
\DeclareBoldMathCommand{\bA}{A}
\DeclareBoldMathCommand{\bJ}{J}

\newcommand\Alfven{Alfv\'en }
\newcommand\Alfvenic{Alfv\'enic }

\newcommand{\V}[1]{\mathbf{#1}} 
\newcommand{\zhat}{\mbox{$\hat{\mathbf{z}}$}}

\begin{document}
\title{An Oscillating Langevin Antenna for Driving Plasma Turbulence Simulations}
\author{J.~M. TenBarge}
\ead{jtenbarg@umd.edu}
\address{IREAP, University of Maryland, College Park, MD, USA}
\author{G.~G. Howes}
\address{Department of Physics and Astronomy, University of Iowa, Iowa City, IA, USA}
\author{W. Dorland}
\address{IREAP, University of Maryland, College Park, MD, USA}
\author{G.~W. Hammett}
\address{Princeton Plasma Physics Laboratory, Princeton, NJ, USA}
\date{\today}
\begin{abstract}
A unique method of driving \Alfvenic turbulence via an oscillating Langevin antenna is presented. This method of driving is motivated by a desire to inject energy into a finite domain numerical simulation in a manner that models the nonlinear transfer of energy from fluctuations in the turbulent cascade at scales larger than the simulation domain.. The oscillating Langevin antenna is shown to capture the essential features of the larger scale turbulence and efficiently couple to the plasma, generating steady-state turbulence within one characteristic turnaround time. The antenna is also sufficiently flexible to explore both strong and weak regimes of \Alfvenic plasma turbulence.
\end{abstract}

\maketitle
\section{Introduction}


The development of a detailed understanding of plasma turbulence is an
outstanding goal of the plasma physics community due to its ubiquity
and importance in a variety of environments. In space physics and
astrophysics, turbulence mediates the transfer of energy from the
large scales at which energy is injected into turbulent motions to the
small scales at which the turbulent energy is ultimately dissipated as
heat. The resulting heating of the plasma determines the radiation
emitted from turbulent astrophysical environments, which constitutes
the majority of our observational data. In the heliosphere, turbulence
likely plays a key role in the heating of the solar corona and in the
launching of the solar wind. The dissipation of turbulent fluctuations
in the streaming solar wind plasma impacts the overall thermodynamic
balance of the heliosphere.

The near-Earth solar wind is a unique laboratory for the study of
plasma turbulence due to its accessibility to direct spacecraft measurements.  The
\Alfvenic nature of the turbulent fluctuations in the solar wind
plasma has long been recognized
\cite{Belcher:1971,Tu:1995,Bruno:2005}.  Modern theories of anisotropic 
MHD turbulence suggest that the physical mechanism that drives the
turbulent cascade of energy from large to small scales is the
nonlinear interaction between counterpropagating \Alfven waves
\cite{Iroshnikov:1963,Kraichnan:1965,Shebalin:1983,Sridhar:1994,
Goldreich:1995,Boldyrev:2006,Howes:2013a}. Although spacecraft missions enable
detailed \textit{in situ} measurements of many aspects of the turbulent plasma
and electromagnetic fluctuations, measurements are generally possible
at only a single point, or at most a few points, in space. The solar
wind plasma typically streams past the spacecraft at super-\Alfvenic
velocities, so a time series of single-point measurements maps to the
advection of spatial variations in the turbulent plasma
\cite{Taylor:1938}. These limitations of spacecraft measurements motivate 
complementary efforts to gain further insight into the nature of
\Alfvenic turbulence using terrestrial laboratory experiments or numerical
simulations. Although the experimental measurement of the nonlinear
interaction between counterpropagating \Alfven waves has recently been
accomplished in the laboratory \cite{Howes:2012b}, the large length
scales and low frequencies associated with \Alfvenic fluctuations are
particularly challenging to realize in the laboratory
\cite{Gekelman:1999}, and experiments thus far have been
limited to the weak turbulence regime \cite{Howes:2012b}. Numerical simulations of plasma turbulence, therefore, are indispensable tools to explore the
fundamental nature of plasma turbulence, the mechanisms of its
dissipation, and the resulting plasma heating.

The simulation of a turbulent plasma system typically requires the injection of
energy into the turbulence at large scale and the dissipation of the
turbulent energy at small scales. For many space and astrophysical
plasma systems of interest, the dynamic range between the observed
energy injection and dissipation scales exceeds current computational
capabilities (a limit of approximately 3 orders of magnitude for 3D
turbulence simulations).  In addition, on the small scales at which
the dissipation mechanisms serve to terminate the turbulent cascade of
energy, the plasma dynamics is weakly collisional in many space and
astrophysical plasmas of interest, so the dissipation is thought to be
governed by some kinetic damping mechanism, such as collisionless
wave-particle interactions
\cite{Howes:2008b,Howes:2008c,Schekochihin:2009}.
Since it is not possible to include, in a single simulation with
realistic physical parameters, both the large-scale process driving
the turbulence and the kinetic physical dynamics governing the
dissipation at small scales, a promising strategy is to focus on a
sub-range of the complete turbulent cascade. 

An exciting frontier in the study of plasma turbulence, one that has
engendered vigorous recent activity \cite{Parashar:2009,Osman:2011,Servidio:2012,TenBarge:2013a,Karimabadi:2013}, is the quest to identify the
physical mechanisms that govern the dissipation of turbulence under
weakly collisional conditions and to determine the resulting heating
of the plasma species. The sub-range of numerical simulations, in this
case, begins with a domain scale that falls within the inertial range
of the turbulent cascade and extends down to encompass the small,
dissipative scales.  Therefore, it is desirable to inject energy into
the simulation at the domain scale in a manner that resembles the
nonlinear transfer of energy, within the inertial range, from scales
slightly larger than the simulation domain. In constructing such a
technique for driving plasma turbulence simulations, it is essential
to account for the inherent scale-dependent anisotropy of
\Alfvenic turbulence \cite{Sridhar:1994,Goldreich:1995,Boldyrev:2006},
which becomes more anisotropic as the turbulence cascades to smaller
scales. Here we describe such a mechanism for forcing plasma turbulence
simulations, the oscillating Langevin antenna, that models the
\Alfvenic fluctuations at the domain scale generated by the 
transfer of energy caused by nonlinear interactions between
counterpropagating \Alfven waves at scales larger than the simulation
domain. The method is effective in generating strong \Alfvenic
turbulence in kinetic simulations and flexible enough to simulate
strong or weak turbulence.

The paper is organized as follows: In \S \ref{sec:turb}, we
introduce some of the basic concepts underlying \Alfvenic
turbulence. \S \ref{sec:drive} discusses the simple case of sinusoidal
driving and plasma coupling before moving on to the more complicated
oscillating Langevin antenna in \S \ref{sec:ola}. In the
latter section, the antenna is described in detail and its domain of
applicability is examined. The implementation of the antenna in
AstroGK is explored and the amplitude necessary for driving strong
turbulence is given in \S
\ref{sec:imp}. \S \ref{sec:compare} briefly discusses driving
methods employed in other turbulence simulations. In \S
\ref{sec:summary}, we present a summary of this paper.

\section{Properties of Turbulence Relevant to Driving Mechanisms}\label{sec:turb}

\begin{figure}[t]
 \includegraphics[width=\linewidth]{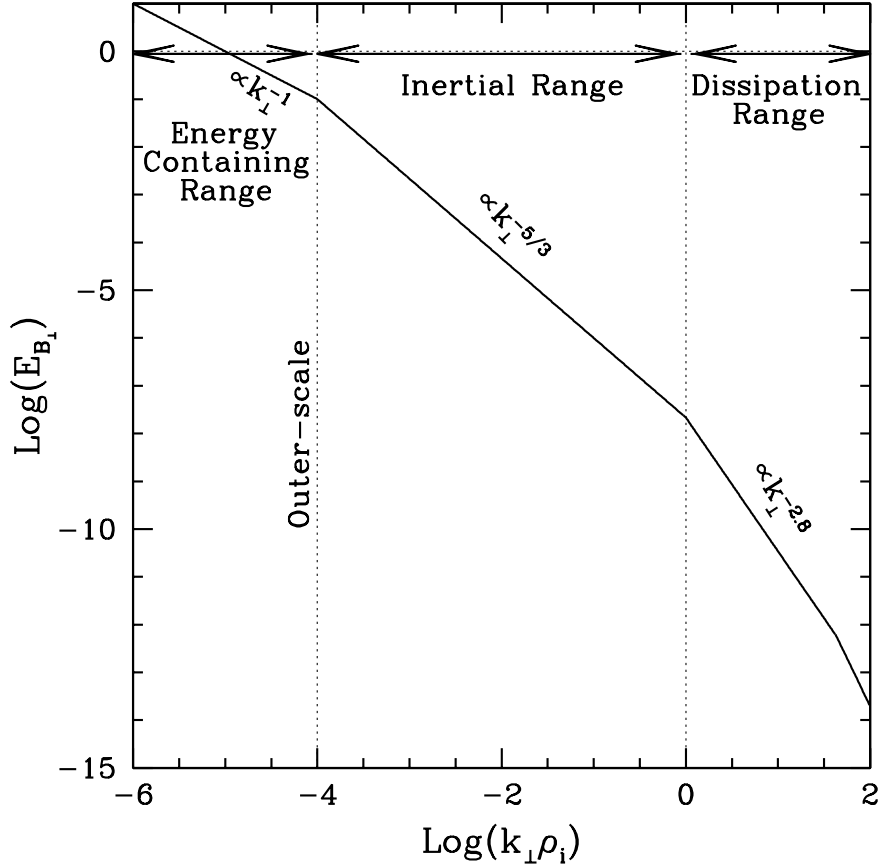} 
\caption{Schematic diagram of the turbulent evolution of energy 
from large to small scales.}\label{fig:Emodel}
\end{figure}

A general picture of the turbulent energy spectrum in weakly
collisional plasma turbulence is shown in Figure~\ref{fig:Emodel},
where the values of the characteristic length scales are appropriate
for the case of turbulence measured in the near-Earth solar wind
\cite{Coleman:1968,Belcher:1971,Leamon:1998a,Howes:2008b,Sahraoui:2009,Kiyani:2009,Alexandrova:2009,Chen:2010,Sahraoui:2010b,Alexandrova:2012}. The transformation from the spacecraft-frame frequency of solar wind measurements to the perpendicular component of the wavevector $k_\perp$ is accomplished by adopting Taylor's hypothesis \cite{Taylor:1938} and by assuming an anisotropy of the turbulent fluctuations, $k_\parallel \ll k_\perp$, as discussed below.  Note that, unless otherwise stated, the energy
discussed throughout will be the one-dimensional perpendicular
magnetic energy as a function of perpendicular wavenumber,
$E_{B_\perp}(k_\perp) =
\int_{-\infty}^\infty \,dk_\parallel\
\int_0^{2\pi} \,d\theta\ k_\perp |\delta
B_\perp(k_\perp,\theta,k_\parallel)|^2 \simeq \delta B_\perp^2 /
k_\perp$, where perpendicular is normal to the local mean magnetic field
and $\delta \bB$ is the fluctuating magnetic field.  The  perpendicular wavenumber
is normalized to the  ion (proton) gyroradius, $\rho_i = v_{ti} /
\Omega_i$, where $v_{ti} = \sqrt{2 T_i / m_i}$ is the ion thermal velocity
and $\Omega_i = e B_0 / m_i c$ is the ion gyrofrequency.

At the largest scales ($k_\perp \rho_i < 10^{-4}$) in
Figure~\ref{fig:Emodel}, the timescale of nonlinear turbulent energy
transfer exceeds the travel time from the sun \cite{Howes:2008b,Wicks:2013}, so
these large scale fluctuations have not yet evolved into a turbulent
cascade. The turbulent fluctuations at these large scales, commonly
denoted the \textit{energy containing range}, are often assumed to be
relatively isotropic with $k_\parallel \sim k_\perp$, and the energy
spectrum is characterized by a spectral index of $-1$.

At the outer scale of the turbulence inertial range ($k_\perp \rho_i
\sim 10^{-4}$) in Figure~\ref{fig:Emodel}, the timescale of the 
nonlinear energy transfer is approximately equal to the travel time of
the solar wind plasma from the sun to 1~AU \cite{Howes:2008b}, so the
turbulent cascade has just had time to develop at this scale, and the
spectrum steepens to a spectral index of $-5/3$. The fluctuations at
the outer scale are assumed to be isotropic, $k_\parallel \sim
k_\perp$. The \textit{inertial range} of the solar wind turbulent energy
spectrum extends from the outer scale of the turbulence down to the
scale of the ion gyroradius, $10^{-4} \lesssim k_\perp \rho_i \lesssim
1$. Within the inertial range, the energy transfer is believed to be
dominated by local interactions in scale, leading to turbulent
dynamics that are self-similar and independent of the driving and the
dissipative microphysics \cite{Schekochihin:2009}.  Theoretical
considerations suggest that the transfer of energy in
\Alfvenic turbulence is anisotropic, where energy is transferred more
effectively to small perpendicular scales than to small parallel scales.
The anisotropy of the turbulence is scale-dependent, with $k_\parallel
\propto k_\perp ^{q}$, with predicted values of $q=2/3$ \cite{Goldreich:1995}
or $q=1/2$ \cite{Boldyrev:2006}. Support for these forms of spectral anisotropy have been observed extensively in the solar wind \cite{Horbury:2008,Podesta:2009a,Wicks:2010a,Luo:2010,Sahraoui:2010b,Chen:2011,Narita:2011} and in numerical simulations \cite{Cho:2000,Maron:2001,Cho:2002a,Mason:2006,Mason:2011,Perez:2012}. Regardless of which scaling
is chosen for the anisotropy, at the end of the inertial range
($k_\perp \rho_i \sim 1$), the turbulent fluctuations are
significantly anisotropic with $k_\parallel \ll k_\perp$. 

The anisotropic turbulent cascade of \Alfven waves in the inertial
range transitions to a cascade of kinetic \Alfven waves at the
perpendicular scale of the ion gyroradius, $k_\perp \rho_i \sim 1$
\cite{Howes:2008a,Howes:2008b,Howes:2008c,Sahraoui:2009,Schekochihin:2009,Sahraoui:2010b,Salem:2012,TenBarge:2012d}. In the
range below this scale ($k_\perp \rho_i \gtrsim 1$), often denoted the
\textit{dissipation range}, the energy spectrum steepens to a spectral
index of approximately $-2.8$, as measured in the solar wind
\cite{Sahraoui:2009,Kiyani:2009,Alexandrova:2009,Chen:2010,Sahraoui:2010b,Alexandrova:2012}
and obtained in kinetic numerical simulations
\cite{Howes:2011b,TenBarge:2012c}. The turbulent energy transfer is expected to become 
yet more anisotropic in this range, $k_\parallel\propto k_\perp^{1/3}$
\cite{Cho:2004,Howes:2008b,Schekochihin:2009}, a prediction supported by
electron MHD \citep{Cho:2004,Cho:2009} and kinetic numerical turbulence
simulations \cite{TenBarge:2012a}.  Within this region, kinetic
damping mechanisms, such as collisionless wave-particle interactions,
can remove energy from the turbulent cascade, ultimately terminating
the cascade and mediating the conversion of turbulent fluctuation
energy to plasma heat.

Note that the scale-dependent anisotropy of the turbulent dynamics has
important implications for the driving of turbulence simulations
within the inertial range wherein $\delta B / B_0 \ll 1$---the waves driven at the simulation domain scale must be anisotropic, at least with $k_\parallel < k_\perp$, and
possibly with $k_\parallel \ll k_\perp$. It is an important, but often
unappreciated, fact that using an isotropic driving mechanism to
simulate a turbulent cascade starting within the inertial range is
inconsistent with the known properties of plasma turbulence \cite{Mason:2006,Perez:2008,Perez:2010a}.

Given the current limitations of modern supercomputers, it is not
possible to simulate self-consistently turbulence from the large
driving scales and follow its evolution down to the smallest scales at
which the turbulent cascade is terminated by some physical dissipation
mechanism.  For example, as depicted in Figure~\ref{fig:Emodel}, the
turbulent cascade begins at the outer scale $k_\perp \rho_i \sim
10^{-4}$, and continues through a transition at the scale of the ion
gyroradius $k_\perp \rho_i \sim 1$ down to the scale of order the
electron gyroradius at $k_\perp \rho_i \sim 10^{2}$.  Modeling the
turbulent cascade in three spatial dimensions over a dynamic range of
6 orders of magnitude while resolving the kinetic physics inherent to
the dissipation mechanism is well beyond computing power for the
foreseeable future. In light of this limitation, a number of different
strategies have been adopted to realize progress on the computational
front: (\#1) simulating turbulence in reduced dimensionality, (\#2)
employing reduced dimensionless ratios of plasma parameters, or (\#3)
modeling turbulence only over a subrange of the turbulent cascade.  We
briefly discuss the limitations of each of these approaches, and then
motivate the development of a physically realistic driving mechanism
for the simulation of a subrange of the turbulent cascade.

Many recent efforts in the simulation of plasma turbulence have
adopted strategy \#1, to simulate the turbulent dynamics in reduced
spatial dimensions
\citep{Gary:2010,Parashar:2009,Markovskii:2011,Vasquez:2012,Donato:2012,Servidio:2012,Wan:2012a,Karimabadi:2013}. The nature of \Alfvenic turbulence, however, is 
inherently three-dimensional, with variation in the direction parallel
to the magnetic field required to describe the physics of \Alfven
waves, and with variation in both dimensions perpendicular to the
magnetic field required to capture the dominant nonlinear term
\cite{Howes:2011b}.  Therefore, any two-dimensional description 
incompletely describes the physical effects that play a role in 
\Alfvenic plasma turbulence.

Strategy \#2 is to employ reduced ratios of plasma parameters to
enable a computationally feasible calculation.  For example, for a
fully ionized hydrogenic plasma of protons and electrons with equal
temperatures $ T_i/T_e=1$, the ratio of the ion to electron
gyroradius $\rho_i/\rho_e$ is equal to the square root of the mass
ratio $m_i/m_e$, which has a physical value $m_i/m_e=1836$. Therefore,
the ratio of gyroradii is $\rho_i/\rho_e= \sqrt {m_i/m_e} \simeq 43$.
Reducing the mass ratio to $m_i/m_e=100$ preserves the limit $m_i/m_e
\gg 1$ but reduces the ratio of gyroradii to $\rho_i/\rho_e= 10$,
lowering the computational resolution required to cover the range of
scales from the ion to the electron gyroradius. The primary
disadvantage of using dimensionless ratios that are reduced from their
physical values is that the numerical results often cannot be directly
compared with observational or experimental measurements.  In
addition, reducing the extent of a finite dynamic range, as in the
reduction of $\rho_i/\rho_e$ from 43 to 10, may lead to not only
quantitative, but possibly also qualitative, changes in the
results. Continuing the example above of using a reduced mass ratio to
reduce the scale separation between $\rho_e$ and $\rho_i$, since
collisionless Landau damping by ions peaks at $k_\perp
\rho_i \sim 1$ and by electrons peaks at  $k_\perp \rho_e \sim 1$, reduction 
of the mass ratio can artificially enhance electron damping at the ion
scale, qualitatively changing the amount of dissipated energy
absorbed by electrons, and likely altering the scaling of the magnetic
energy spectrum in the dissipation range, since the spectrum is
determined by a balance of the rate of nonlinear energy transfer to
the rate of collisionless damping \cite{Howes:2008b,Howes:2011b}.

The viewpoint of the authors is that strategy \#3, to model the
turbulence only over a subrange of the turbulent cascade, can achieve
a computationally feasible calculation without sacrificing the
advantage of physical realism. Since a steady-state turbulent spectrum
can only be achieved by a steady rate of energy injection into the
turbulence at large scales and an equal rate of resolved turbulent
energy dissipation at small scales, this problem reduces to
constructing appropriate numerical mechanisms for turbulent driving
and dissipation. Turbulence simulations that directly model the
driving scale of the turbulence require some numerical dissipation
mechanism to remove energy from the simulation at the smallest
resolved scales (where the scale corresponding to the physical
dissipation mechanism is unresolved\footnote{Historically, simulations
of MHD turbulence have followed this strategy, modeling the driving of
the turbulence at the outer scale and using viscosity and resistivity
as the dissipation mechanism. Viscosity and resistivity are fluid
closures for the dissipation in the strongly collisional limit, a
limit not applicable to the weakly collisional conditions of many
space, astrophysical, and laboratory plasmas of interest.  Therefore,
viscosity and resistivity in MHD turbulence simulations often
represent \textit{ad hoc} numerical mechanisms for dissipation,
generally not rigorous models of the physical mechanisms responsible
for the dissipation of turbulence in a weakly collisional plasma.}).
Similarly, simulations that aim to model directly the physical
dissipation mechanisms require some numerical mechanism for injecting
energy into the turbulence at the domain scale (where the scale
corresponding to the physical driving mechanism is larger than the
simulation domain). It is also possible to model the central range of
the turbulent cascade (for example, to model the transition at
$k_\perp
\rho_i \sim 1$ in Figure~\ref{fig:Emodel}), where both a numerical
driving mechanism and a numerical dissipation mechanism are
required \cite{Howes:2008a}. \emph{The focus of this paper is to describe a physically
motivated mechanism for numerical driving of plasma turbulence simulations
where the simulation domain is smaller than the physical driving scale
of the turbulent cascade. {The key physical concept motivating our implementation of the driving is that the turbulent cascade is mediated by nonlinear interactions between counterpropagating Alfv\'{e}n waves. We find that this manner of driving minimizes the transition from the driving scale to the inertial regime of the turbulence, thereby maximizing the effective dynamic range of the turbulent cascade.}}

\begin{figure}[t]
		\includegraphics[width=\linewidth]{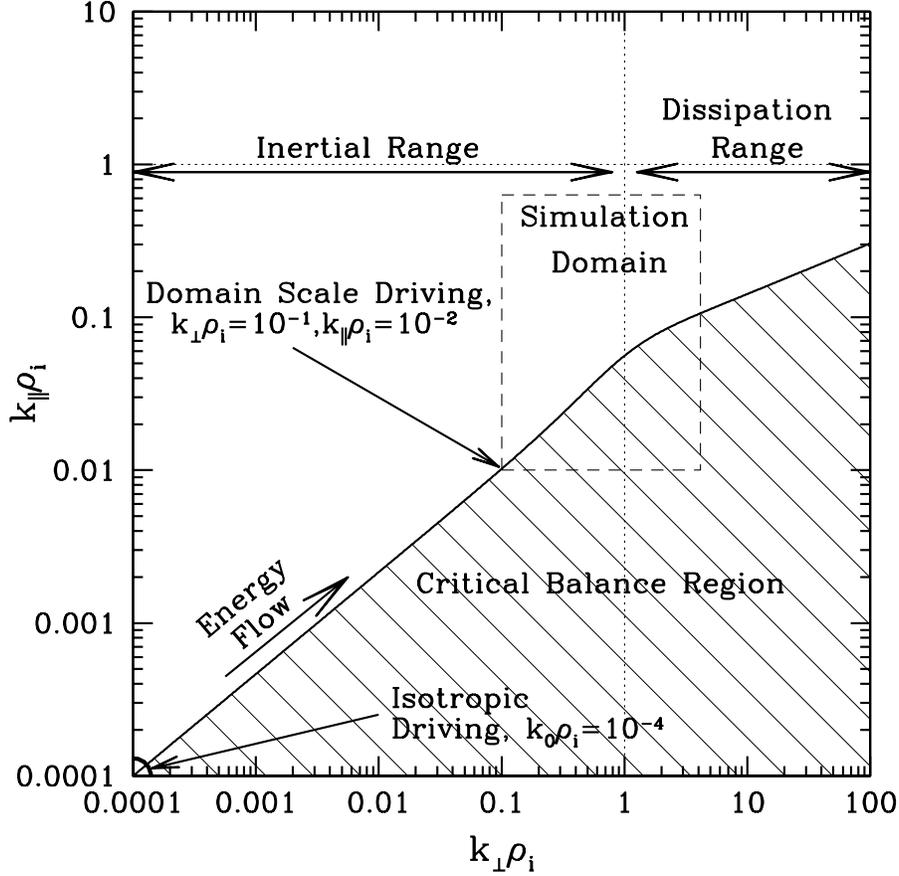}
		\caption{Schematic diagram of the distribution of
		energy in the $k_\perp - k_\parallel$~plane,
		highlighting the turbulence driving scale, typical
		simulation domain, and domain scale
		driving.}\label{fig:driving}
\end{figure}

The effort to devise an appropriate numerical mechanism for driving
turbulence to simulate the small-scale end of the turbulent cascade is
motivated by the intense interest that has recently arisen in
identifying and characterizing the physical processes responsible for
the dissipation of plasma turbulence and the consequent heating of the
plasma species.  Vigorous efforts to address this new frontier in
plasma turbulence have been driven by significant advances in
observational studies, enabling direct measurements of the turbulence
at the dissipative scales in the near-Earth solar wind
\cite{Sahraoui:2009,Kiyani:2009,Alexandrova:2009,Chen:2010,Sahraoui:2010b,Alexandrova:2012},
and in numerical methods for study of kinetic dissipation mechanisms
\cite{Howes:2008a,Saito:2008,Parashar:2009,Numata:2010,Saito:2010,Parashar:2010,Gary:2010,Howes:2011b,Chang:2011,TenBarge:2012a,Gary:2012,Servidio:2012,TenBarge:2013a,Karimabadi:2013}.

The anisotropy inherent in magnetized plasma turbulence introduces
subtle, but important, complications in devising an appropriate
mechanism for turbulent energy injection into a simulation domain
whose largest scale falls within the inertial range. The
characteristic anisotropy in wavevector space over different scale
ranges of \Alfvenic plasma turbulence, as described earlier in this
section, is depicted in Figure~\ref{fig:driving}. At the outer scale
$k_\perp \rho_i \sim 10^{-4}$, the turbulent fluctuations are believed
to be approximately isotropic, with $k_\parallel \sim k_\perp$. But
energy is transferred through the inertial range preferentially to smaller
perpendicular scales, as described by the scale-dependent anisotropy,
$k_\parallel
\propto k_\perp ^{q}$, where we have adopted the Goldreich-Sridhar 
model for turbulence with  $q=2/3$ \cite{Goldreich:1995} 
to construct this figure. When the turbulent cascade enters the
dissipation range at $k_\perp \rho_i \sim 1$, the turbulence becomes
yet more anisotropic, with a scaling $k_\parallel\propto
k_\perp^{1/3}$
\cite{Cho:2004,Howes:2008b,Schekochihin:2009,TenBarge:2012a}. Turbulent power is
believed to fill the region below this boundary of critical balance,
filling the shaded region in the figure
\cite{Goldreich:1995,Maron:2001,Howes:2008b,TenBarge:2012a}. A simulation 
intended to model the central part of the turbulent cascade, as
depicted in Figure~\ref{fig:driving}, must take into account the
characteristic anisotropy of the turbulent fluctuations in three
important ways. First, the simulation domain must reflect the
anisotropy of the turbulence, which motivates using an elongated
domain to describe optimally the turbulent fluctuations. And, second,
the numerical mechanism for driving the turbulence must drive
fluctuations with the expected anisotropy of the turbulence at the
domain scale. The third and final consideration, if the simulation is
intended to model the case of strong turbulence
\cite{Sridhar:1994,Goldreich:1995}, is that the amplitude of the 
turbulent driving should generate turbulent fluctuations satisfying
the condition of critical balance \cite{Higdon:1984a,Goldreich:1995}. {These three aspects of simulating turbulence have been implemented by a number of other authors, e.g., \cite{Maron:2001,Perez:2008,Beresnyak:2009,Perez:2010a}.}

The wavenumber range in $(k_\perp,k_\parallel)$ that is covered by a
modest simulation of a subrange of the physical turbulent cascade is
depicted in Figure~\ref{fig:driving} by the dashed box.  Note that, in
$(k_\perp,k_\parallel)$ space, the lower left corner of this dashed
box corresponds to the perpendicular and parallel domain scale, and
the upper right corner corresponds to the perpendicular and parallel
resolution. In the case plotted in Figure~\ref{fig:driving}, the
simulation domain scale is given by $(k_{\perp 0} \rho_i,k_{\parallel
0} \rho_i)= (10^{-1},10^{-2})$, so that the simulation domain is an
elongated box of dimensions $L_{\parallel 0} \times L_{\perp 0}^2$,
where $L_{\parallel 0}=200 \pi \rho_i$ and $L_{\perp 0}=20 \pi
\rho_i$. Similarly, the fully de-aliased resolution of the simulation is 
$\Delta l_\parallel \simeq 4.5 \pi \rho_i$ and $\Delta l_\perp\simeq \pi
\rho_i/2 $. The elongation of the simulation domain is chosen so that 
the largest scale perpendicular and parallel fluctuations correspond
to the anisotropic fluctuations that coincide with the condition of
critical balance (the line along the upper boundary of the shaded
region in Figure~\ref{fig:driving}). By driving the anisotropic
turbulent fluctuations at the domain scale (in which $k_{\perp 0} >
k_{\parallel 0}$) at the amplitude specified for critical balance, a
cascade of strong plasma turbulence can be driven. This numerical
mechanism for driving the turbulence is inspired by the physical
properties of \Alfvenic turbulence.

Modern theories of anisotropic incompressible MHD turbulence are based
on the key concept that the turbulent cascade of energy from large to
small scales is driven by the nonlinear interaction between
counterpropagating \Alfven waves
\cite{Kraichnan:1965,Sridhar:1994,Goldreich:1995,Boldyrev:2006,Howes:2012b}. 
Therefore, our numerical mechanism for driving the turbulence in a
physically realistic manner is to drive \Alfven waves that travel both
up and down the equilibrium magnetic field. Since the \Alfven wave in
the MHD limit, $k\rho_i \ll 1$, has no parallel magnetic field
perturbation, $\delta B_\parallel = 0$, an
\Alfven wave can be driven effectively by applying a  parallel current
throughout the simulation domain (corresponding to generating a body
force).  The parallel current is applied across the domain as a plane
wave with the wavevector and frequency of the desired \Alfven wave.
If the driven waves reach amplitudes satisfying the condition of
critical balance, nonlinear interactions between these driven
\Alfven waves rapidly generate a turbulent cascade of energy to small
scales. The condition of critical balance implies that the timescale
of the nonlinear energy transfer balances the linear timescale of the
interacting \Alfven waves, thus the frequency of the plane wave component
of parallel current is chosen to be approximately the linear frequency
of the domain scale \Alfven waves, and the driving should decorrelate
on approximately the same timescale. The aim of this numerical driving
method is to model the nonlinear transfer of energy from
counterpropagating \Alfven waves at scales larger than the simulation
domain to the domain scale \Alfven waves. This physically motivated
driving is accomplished with the \textit{Oscillating Langevin Antenna},
defined and characterized in the remainder of this paper.

\section{A Simple Model of Antenna Driving}\label{sec:drive}

Before discussing the complexities of an antenna driven by a Langevin
equation, we begin by discussing the simple case of sinusoidal driving
of an incompressible MHD plasma to understand the general response of the
plasma as a function of driving frequency.  The incompressible MHD
equations can be cast in Elsasser form \cite{Elsasser:1950} as
\begin{equation}
\frac{\partial \V{z}^{\pm}}{\partial t} 
\mp \V{v}_A \cdot \nabla \V{z}^{\pm}
=-  \V{z}^{\mp}\cdot \nabla \V{z}^{\pm} -\nabla P/\rho_0  + \nu \nabla^2 \bz^\pm + \bF^\pm,
\label{eq:elsasserpm}
\end{equation}
\begin{equation}
\nabla\cdot  \V{z}^{\pm}=0,
\label{eq:div0}
\end{equation}
where the magnetic field is decomposed into equilibrium and
fluctuating parts $\V{B}=\V{B}_0+ \delta
\V{B} $, $\V{v}_A =\V{B}_0/\sqrt{\mu_0\rho_0}$ is the \Alfven velocity 
due to the equilibrium field $\V{B}_0=B_0 \zhat$, $P$ is total
pressure (thermal plus magnetic), $\rho_0$ is mass density, the
Laplacian term leads to damping where viscosity and resistivity have
been set equal $\nu=\eta$, $\bF^\pm$ is an external forcing, and
$\V{z}^{\pm} =\V{u} \pm \delta \V{B}/\sqrt{4 \pi \rho_0}$ are the
Els\"asser fields given by the sum and difference of the velocity
fluctuation $\V{u}$ and the magnetic field fluctuation $\delta \V{B}$
expressed in velocity units. The divergence condition \eqref{eq:div0}
specifies the pressure in the incompressible plasma
\cite{Howes:2013a},
\begin{equation}
\nabla^2 P/\rho_0 =  - \nabla \cdot \left(   \V{z}^{\mp}\cdot \nabla \V{z}^{\pm}\right).
\label{eq:press}
\end{equation}
  The second term on the left-hand side of
\eqref{eq:elsasserpm} is the linear term representing the propagation
of the Els\"asser fields along the mean magnetic field at the \Alfven
speed, the first term on the right-hand side is the nonlinear term
representing the interaction between counterpropagating waves, and the
second term on the right-hand side is a nonlinear term that enforces
incompressibility through \eqref{eq:press}. 

Focusing on the evolution of a single Fourier mode $\bz_k^\pm$, we
choose to write \eqref{eq:elsasserpm} in the simplified functional
form
\begin{equation}\label{eq:els2}
\frac{\partial \bz_k^\pm}{\partial t} \mp i \omega_l \bz_k^\pm + \gamma \bz_k^\pm  = \bF_k^\pm,
\end{equation}
where the second term on the left-hand side represents the
characteristic linear response where the linear frequency is
$\omega_l=k_\parallel v_A$, and the third-term represents the loss of
energy from the single Fourier mode $\bz_k^\pm$ with effective
frequency $\gamma$.  Here, the loss term $\gamma =
\omega_{nl} + \nu k^2$ combines two separate physical  effects:
(i) linear damping due to viscosity and resistivity given by $\nu
k^2$; and (ii) the nonlinear transfer of energy through interactions
with all other Fourier modes via the nonlinear interaction and pressure
terms, where we generalize the effects of these terms as a nonlinear
frequency response, $\omega_{nl} \bz_k^\pm
\sim \bz^\mp \cdot \nabla \bz_k^\pm + \nabla P/\rho_0$.

The physics described by \eqref{eq:els2} is similar to the physics of
the driven, damped harmonic oscillator, described by
\begin{equation}
\frac{d^2 x}{d t^2} + \gamma \frac{d x}{dt} + \omega_l^2 x  = F.
\label{eq:ddsho}
\end{equation} 
If $F$ is assumed to be a sinusoidal driving of the form $F = A_0
\sin{\omega t}$, then the steady-state portion of the solution to 
\eqref{eq:ddsho} is
\begin{equation}
x(t) = \frac{A_0}{\sqrt{(\omega_l^2-\omega^2)^2 + \gamma^2 \omega^2}}
\sin{(\omega t + \delta)},
\end{equation}
where $\delta = \arctan{\left[-\omega \gamma / (\omega_l^2 -
\omega^2)\right]}$.  
Solving \eqref{eq:els2} via a similar Fourier transform method yields
a  steady-state amplitude response of
\begin{equation}
|\bz_k^\pm| = \frac{A_0}{\sqrt{(\omega_l-\omega)^2 + \gamma^2}}.
\end{equation}
Although the mathematical forms of these solutions to 
\eqref{eq:els2} and \eqref{eq:ddsho} are not identical, they do share 
a similar qualitative form, demonstrating that the solution for a
single Fourier mode in a sinusoidally driven incompressible MHD plasma
will share similar physical properties with the driven, damped
harmonic oscillator.

\begin{figure}[t]
		\includegraphics[width=\linewidth]{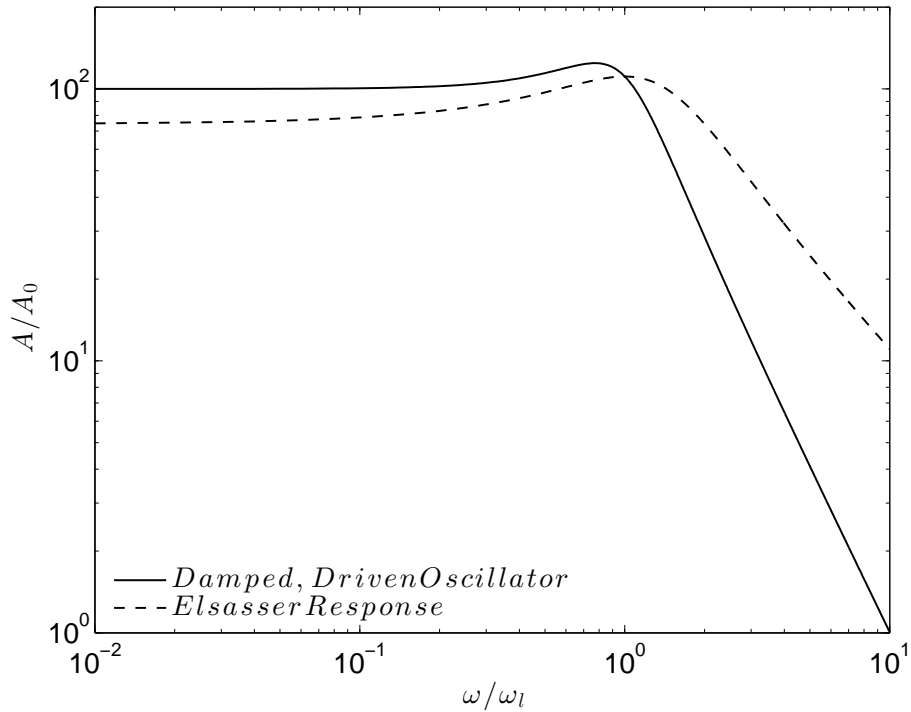}
		\caption{The normalized amplitude $A/A_0$
		vs.~normalized driving frequency $\omega/\omega_l$ for
		a driven, damped harmonic oscillator (solid) and for
		\eqref{eq:els2} describing an incompressible MHD
		plasma (dashed) with $\gamma/\omega_l =
		0.9$.}\label{fig:amp_ddo} 
\end{figure}

We now consider the response of a single Fourier mode in \Alfvenic
turbulence according to \eqref{eq:els2}. The condition of critical
balance in the case of strong MHD turbulence implies a balance between
the linear and nonlinear frequencies $\omega_{nl} \sim \omega_l$
\cite{Goldreich:1995,Howes:2008b,Howes:2011b}. In addition, we are interested in 
the regime where the linear damping rate is weak, $\nu \ll
\omega_l$. Consequently, we expect the energy loss rate for  a single 
Fourier mode in strong \Alfvenic turbulence to be represented by
$\gamma \sim \omega_l$. 

For sinusoidal driving of the form $A_0 \sin{\omega t}$, we plot the
normalized amplitude response $A/A_0$ as a function of the normalized
driving frequency $\omega/\omega_l$ for both the case of the driven,
damped harmonic oscillator (solid) and the case of the solution to
Elsasser equation \eqref{eq:els2} for an incompressible MHD plasma
(dashed) in Figure~\ref{fig:amp_ddo}. In this plot, we have taken the
energy loss rate, dominated by nonlinear energy transfer in the case
of the turbulent plasma, as $\gamma/\omega_l =0.9$. Defining the
resonant frequency $\omega_R$ as the driving frequency resulting in
the maximum amplitude response, for the case of the driven, damped
harmonic oscillator, we obtain $\omega_R/\omega_l = \sqrt{1 -
(1/2)(\gamma/\omega_l)^2 }$; for the case of the incompressible MHD
plasma, we obtain $\omega_R/\omega_l=1$. For driving frequencies below
the resonant frequency, $\omega < \omega_R$, the resulting amplitude
is essentially constant. For frequencies above resonance, however, the
coupling becomes very poor, due to the injection and removal of energy
faster than the oscillator or plasma can respond.
 
Using the intuition gained from this exploration of the linear
response of a given mode, we address the issue of how to choose an
appropriate driving frequency for simulations of plasma turbulence.
In order to obtain efficient coupling between the driving antenna and
the plasma, the analysis above demonstrates that one must drive with a
frequency at or below the resonant frequency, which is approximately equal to the
linear frequency. Therefore, we choose a driving frequency $\omega
\lesssim \omega_l$.  If one attempts to drive the plasma with a
frequency above the linear frequency of the wavenumber mode being
excited, there will be an impedance mismatch and very little energy
will enter the plasma, analogous to the case of a damped, driven
harmonic oscillator. This effect has been observed by \citet{Parashar:2011} in 2D hybrid kinetic numerical simulations. For the case of strong turbulence, the nonlinear
energy transfer frequency balances the linear frequency, $\omega_{nl} \sim \omega_l$
\cite{Goldreich:1995,Howes:2008b,Howes:2011b}.  Since our antenna aims to 
model the nonlinear energy transfer from fluctuations at slightly
larger scales than our simulation volume, we expect the driving
frequency to be given by the nonlinear frequency, $\omega
\sim \omega_{nl}$.  Therefore, for realistic driving of turbulence simulations
at a driving scale within the turbulent inertial range, the
appropriate choice is a driving frequency that is the same order of
magnitude as the linear frequency, $\omega \sim \omega_l$, which
fortunately ensures good coupling of the antenna to the turbulent
plasma.

\section{Oscillating Langevin Antenna}\label{sec:ola}

\subsection{Numerical  Implementation of Antenna}
We wish to construct an antenna that will drive \Alfvenic fluctuations
in a simulation of plasma turbulence. For a plasma with an equilibrium
magnetic field, $\bB_0 = B_0 \hat{\bz}$, the eigenfunction for the
\Alfven wave has no magnetic field fluctuation in the direction
parallel to the equilibrium magnetic field, $\delta B_z =
0$. Maxwell's equations require that $\nabla \cdot \bB =0$, which
reduces to $\bk_\perp \cdot \delta \bB_\perp = 0$ for a plane \Alfven
wave with wavevector $\bk= \bk_\perp + k_z \zhat$. To drive a general
perpendicular magnetic field fluctuation $ \delta \bB_\perp $, one can
impose through the plasma a current parallel to the equilibrium
magnetic field, $\bJ=J_z \hat{\bz}$.  Since we can express the
magnetic field fluctuation in terms of the curl of a vector potential,
$\delta \bB = \nabla \times \bA$ and the current in terms of the curl
of the magnetic field, $\bJ = (c/4 \pi) \nabla \times \bB$, a parallel
current can be generated by driving the parallel component of the
vector potential according to $J_z = -(c/4 \pi) \nabla^2 A_z$.  The magnetic field generated by this parallel vector potential is given by
\begin{equation}\label{eq:bperp}
\delta \bB_\perp = - \hat{\bz} \times \nabla A_z =
 i \left(k_y A_z\hat{\bx} - k_x A_z \hat{\by}\right).
\end{equation}
Therefore, a given Fourier mode of $\delta \bB_\perp$ can be 
generated by specifying a  Fourier mode of $A_z$ with wavevector
$\bk= k_x \hat{\bx} +  k_y \hat{\by} + k_z \hat{\bz} $.


Our implementation of the \emph{oscillating Langevin antenna} drives a 
parallel body current through the plasma, where the properties of the 
antenna are specified by four parameters:
\begin{enumerate}
\item Wave vector $\bk$
\item Amplitude $A_0$
\item Characteristic frequency $\omega_0$ (real)
\item Decorrelation rate $\gamma_0 < 0$ (real).
\end{enumerate}
Note that the frequency and decorrelation rate can be written together
more concisely in complex notation as $\omega_a = \omega_0 + i \gamma_0$. The parallel
vector potential of the antenna $A_{za}$ is given by
\begin{equation}
A_{z a}(k_x,k_y,k_z,t) = a_n e^{i \bk \cdot \br},
\end{equation}
where the discrete (complex) value of the driving coefficient at
timestep $n$ is $a_n = a(t_n)$. We initialize the driving coefficient
as $a_0 = A_0
\exp{(i
\phi)}$, where the choice of phase, $\phi$, is arbitrary.

The driving coefficient is evolved by the equation
\begin{equation}\label{eq:lang_code}
a_{n+1} = a_n e^{-i \omega_a \Delta t} + F_a \Delta t,
\end{equation} 
where
\begin{equation}
F_a = \sigma u_n,
\end{equation}
$u_n$ is a delta-correlated uniform complex random number,
$Re(u_n) \in \left[-1/2, 1/2\right]$ and $Im(u_n) \in \left[-1/2, 1/2\right]$, $\sigma$ is an amplitude to be determined by the requirement that the antenna amplitude satisfy $\langle |a_n|^2 \rangle = A_0^2$, and brackets indicate an ensemble average. Applying this requirement to equation~\eqref{eq:lang_code} leads to
\begin{equation}
A_0^2 \left(1 - e^{2 \gamma_0 \Delta t} \right) = \sigma^2 \Delta t^2 \langle |u_n|^2 \rangle.
\end{equation}
For numerical convergence, we require that $|\omega_a| \Delta t \ll 1$. Also, $\langle |u_n|^2 \rangle = 1/6$ for $u_n$ as defined above. Therefore, 
\begin{equation}
\sigma = A_0 \sqrt{12 \frac{|\gamma_0| }{\Delta t}}.
\end{equation}
Note that this definition of $\sigma$ is only valid for $u_n$ as defined above. For instance, $u_n$ constructed from a Gaussian complex random number would result in a different value for $\sigma$.

The relation between \eqref{eq:lang_code} and a standard stochastic
process becomes more apparent when converted to the form of a finite
difference advance in time using $e^{-i \omega_a \Delta t}
\simeq 1 - i \omega_a \Delta t$ to yield
\begin{equation}
\frac{a_{n+1} - a_n}{\Delta t} = -i w_a a_n + F_a.
\end{equation}
The continuous version of this equation is thus
\begin{equation}\label{eq:lang_cont}
\frac{d a}{dt} = -i \omega_0 a + \gamma_0 a + F_a.
\end{equation}
If we take $a = dv / dt$, then Fourier transforming in time yields $a
= -i \omega_0 v$. Substituting this relation into \eqref{eq:lang_cont}
shows that this is the equation of a stochastically driven and damped
harmonic oscillator
\cite{Lemons:2002}
\begin{equation}\label{eq:sdho}
\frac{d a}{dt} =  - \omega_0^2 v + \gamma_0 a + F_a.
\end{equation}

Physically, \eqref{eq:sdho} describes the
motion of a harmonic oscillator with characteristic frequency
$\omega_0$ and damping rate $\gamma_0$, where damping is due to stochastic
particle collisions which serve to both damp the oscillator and cause
random fluctuations. In the absence of the oscillating term in
\eqref{eq:sdho}, it reduces to the standard Langevin equation
describing Brownian motion \cite{Langevin:1908}. Although our antenna
is not governed by particle collisions, it obeys \eqref{eq:sdho};
therefore, we refer to our antenna as an \emph{oscillating Langevin antenna}.

An important note should be made here: Although our driving method
includes a Gaussian random number, $u_n$, which represents white
noise, \eqref{eq:lang_code} integrates the white noise to achieve
Brownian motion. White noise, depicted in Figure~\ref{fig:white},
applies energy equally at all frequencies, while the oscillating
Langevin antenna applies energy with a peak centered on the driving frequency.

Fourier transforming \eqref{eq:lang_cont} yields a squared amplitude
response of the antenna in the form of a Lorentzian peaked about the
antenna driving frequency
\begin{equation}
|a|^2 = \frac{\sigma^2 |\tilde{u}_n|^2}{(\omega - \omega_0)^2 + \gamma_0^2},
\end{equation}
where $\tilde{u}_n$ is the Fourier transform of $u_n$ (white
noise). As expected, the full-width at half-maximum (FWHM) of the
antenna amplitude is proportional to the decorrelation rate, FWHM$= 2
\sqrt{3} \gamma_0$.

\subsection{Examples of Driving by Oscillating Langevin Antenna}

\begin{figure}[t]
		\includegraphics[width=\linewidth]{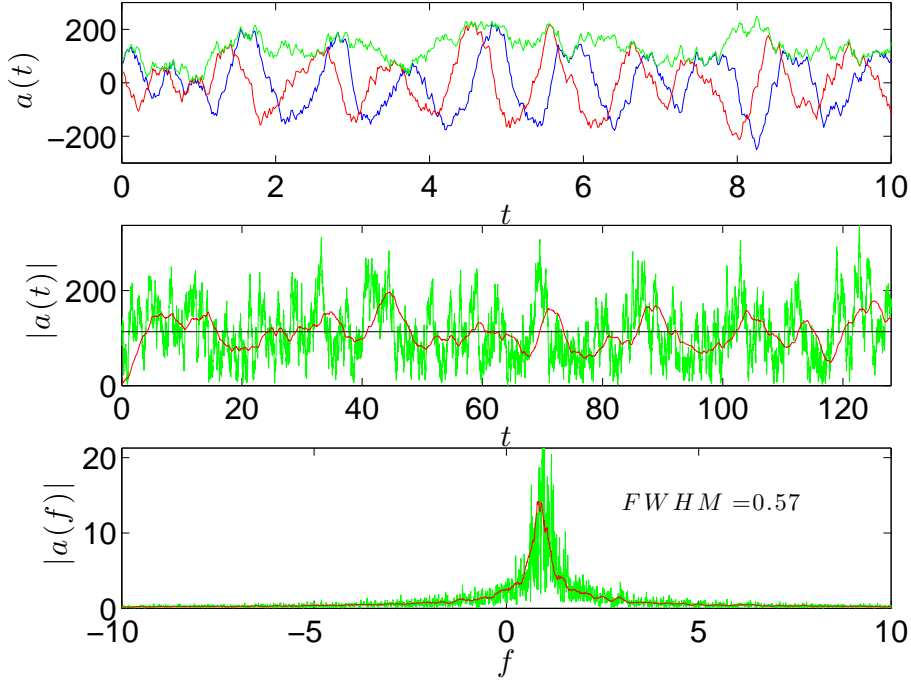}
		\caption{Temporal evolution of a single oscillating
		Langevin antenna with $A_0 = 100$, $\omega_0 = 2 \pi$~rad/s,
		and $\gamma_0 = -1$~rad/s. Top: Real (blue), imaginary
		(red), and the magnitude (green) of the complex vector
		potential over 10 periods. Middle: Magnitude (green),
		boxcar average of the magnitude (red), and overall
		average of the magnitude (black) of $a$ over 128
		periods. Bottom: The frequency spectrum of the antenna
		(green) and its boxcar average (red) along with the
		full-width at half-maximum.}\label{fig:langevin1}
\end{figure}

In Figure~\ref{fig:langevin1}, we present an example of the evolution
and frequency response of the oscillating Langevin antenna with
amplitude $A_0 = 100$, angular frequency $\omega_0 = 2 \pi$~rad/s, and
decorrelation rate $\gamma_0 = -1$~rad/s.  Note that this driving angular
frequency corresponds to a \emph{linear} frequency $f=1$~Hz.  The figure
was produced by evolving equation~\eqref{eq:lang_code} for 128 periods
with $\Delta t = 2
\pi / (\omega_0 n_p) = 1/64$~s, where $n_p = 64$ is the number of points
per period. The Nyquist frequency is thus $\omega_{Nq} = 2 \pi / 2
\Delta t = \omega_0 n_p / 2 = 64 \pi$~rad/s. In the upper panel, the real
(blue), imaginary (red), and amplitude (green) of the antenna are
plotted over only the first ten periods. In the second panel, the amplitude (green),
the amplitude with a boxcar average with width $250$ points applied
(red), and the average amplitude (black) are plotted for the full 128
periods. In the third panel, the Fourier transform of the antenna
(green) and a boxcar averaged (width of $25$ points) version of the
amplitude (red) of the full 128 periods are plotted. Performing an
average of 64 ensembles, we obtain an average amplitude $<|a|> = 103 \pm 54$,  average linear frequency $<f> = 1.003
\pm 0.083$~Hz, and average $<$FWHM$> = 0.57 \pm 0.08$~ Hz, where brackets
indicate ensemble averages and errors represent one standard
deviation. The theoretical FWHM based on the Lorentzian response is
FWHM$= 0.55$~Hz. The ensemble averages and the temporal response of
the antenna represent excellent agreement with the values specified
for the driving.

\begin{figure}[t]
		\includegraphics[width=\linewidth]{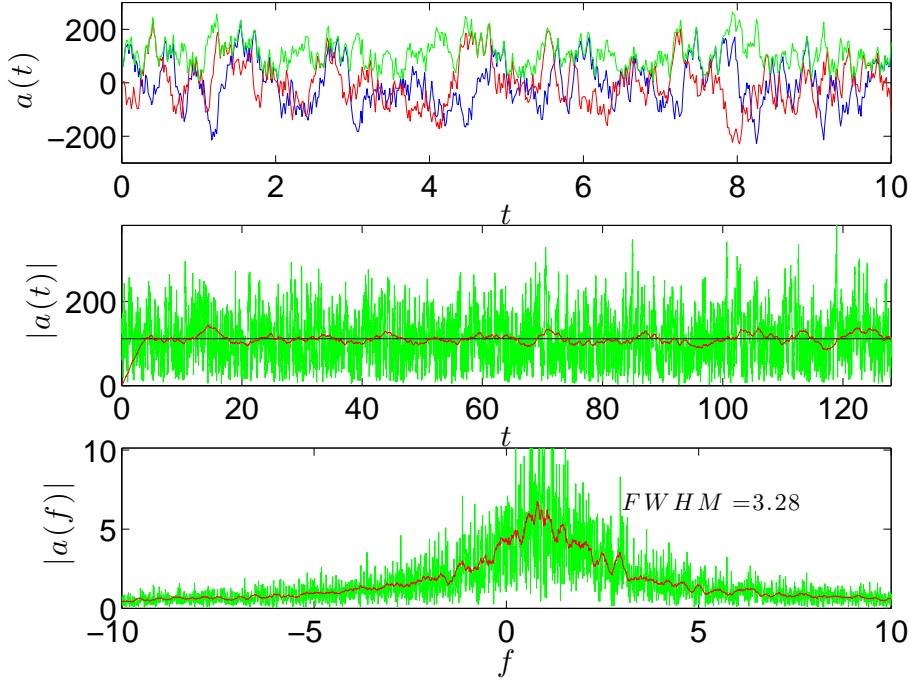}
		\caption{Temporal evolution of a single oscillating
		Langevin antenna with $A_0 = 100$, $\omega_0 = 2 \pi$~rad/s,
		and $\gamma_0 = -2 \pi$~rad/s. Top: Real (blue), imaginary
		(red), and the magnitude (green) of the complex vector
		potential over 10 periods. Middle: Magnitude (green),
		boxcar average of the magnitude (red), and overall
		average of the magnitude (black) of $a$ over 128
		periods. Bottom: The frequency spectrum of the antenna
		(green) and its boxcar average (red) along with the
		full-width at half-maximum.}\label{fig:langevin2}
\end{figure}
A similar single simulation with parameters $A_0 = 100$, $\omega_0 = 2
\pi$~rad/s, $\gamma_0 = -2 \pi$~rad/s, and $n_p = 64$ is presented in 
Figure~\ref{fig:langevin2}. Again, performing an average of 64
ensembles, we find $<|a|> = 107 \pm 55$, $<f> = 0.97 \pm
0.32$~Hz, and $<$FWHM$> = 3.1 \pm 0.3$~Hz. The theoretical FWHM
based on the Lorentzian response is FWHM$ = 3.4$~Hz. As expected
for this case, the amplitude and central frequency are statistically
unchanged from the first simulation, but the decorrelation rate
(quantized by the FWHM) is larger, indicating the energy of the
antenna will be spread into a wider range of frequencies than the
first case.

\subsection{Characterization of Antenna Behavior}

The effective independent dimensionless variables determining the
antenna evolution for a given amplitude $A_0$ and driving frequency
$\omega_0$ are the normalized decorrelation rate $\gamma_0/\omega_0$
and normalized time step $\omega_0 \Delta t$. To explore the region of
validity of the antenna, we choose fiducial values corresponding to
the second set of parameters presented above, $A_0 = 100$, $\omega_0 =
2 \pi$~rad/s, $\gamma_0 = -2\pi$~rad/s, and $\omega_0 \Delta t = 2 \pi
/ n_p = \pi / 32$ and separately vary $\gamma_0$ and $\Delta t$ to
characterize the behavior of the antenna in terms of the resulting
average amplitude, FWHM, and frequency. 

\begin{figure}[t]
		\includegraphics[width=\linewidth]{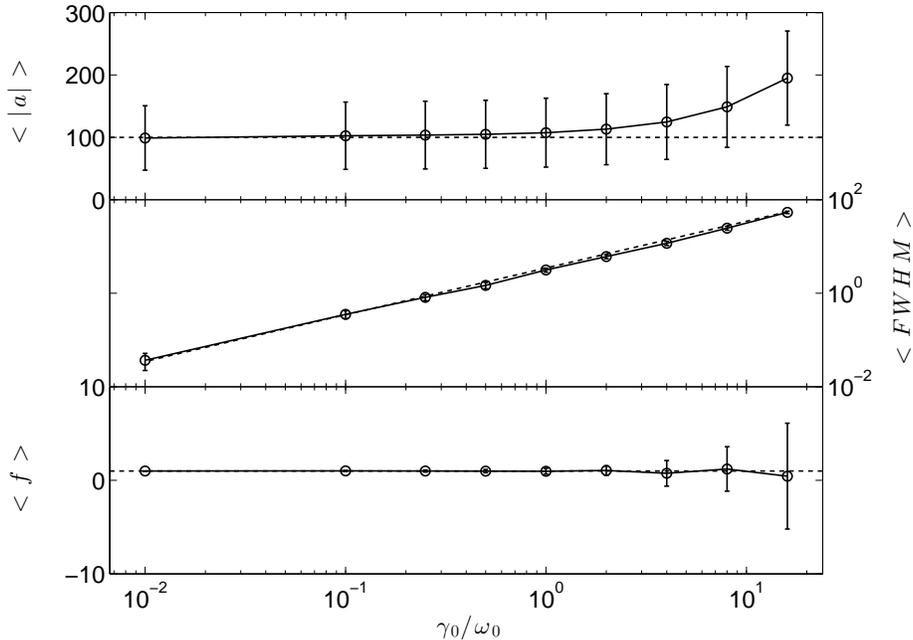}
		\caption{The average amplitude, FWHM, and central
		frequency of the oscillating Langevin antenna over 64
		identically prepared ensembles with $A_0 = 100$,
		$\omega_0 = 2 \pi$~rad/s, and $\omega_0 \Delta t = \pi / 32$
		fixed as the decorrelation rate, $\gamma_0$, is
		varied. The circles and error bars represent the
		ensemble average and standard deviation, and the
		dotted lines represent the theoretical
		values.}\label{fig:vargam}
\end{figure}

The results of varying the decorrelation rate $\gamma_0$ are presented
in Figure~\ref{fig:vargam}. The circles correspond to ensemble
averages over 64 ensembles, error bars are the standard deviation, and
dotted lines correspond to the expected theoretical values. Although
the FWHM is well-behaved for all values of $\gamma_0$, the average
amplitude and central frequency begin to deviate significantly from
their theoretical values for $\gamma_0 /\omega_0 \gtrsim
4$. 

\begin{figure}[t]
		\includegraphics[width=\linewidth]{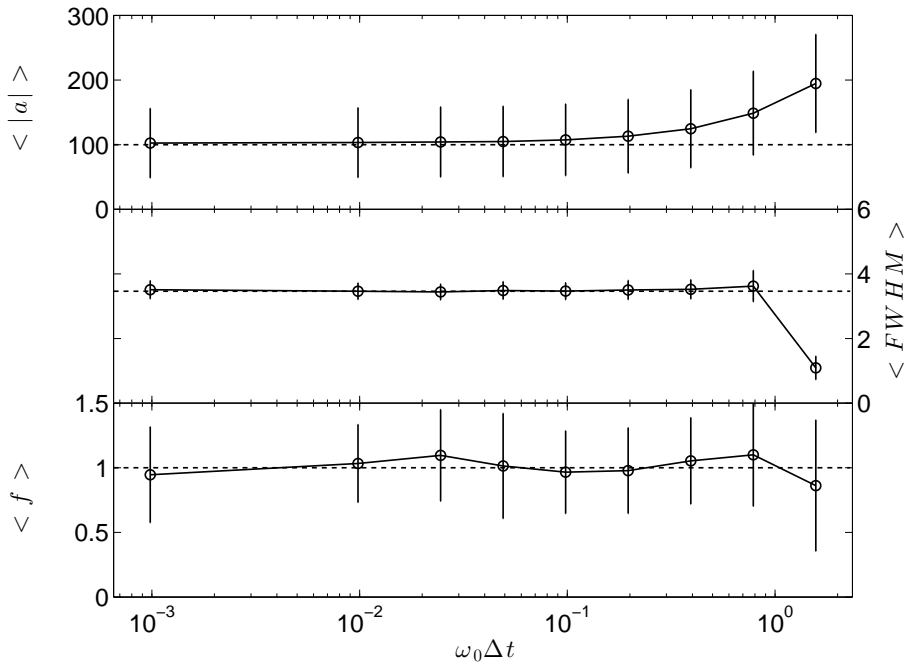}
		\caption{The average amplitude, FWHM, and central
		frequency of the oscillating Langevin antenna over 64
		identically prepared ensembles with $A_0 = 100$ and
		$\omega_0 = -\gamma_0 = 2 \pi$~rad/s fixed as the time
		step, $\Delta t$, is varied. The circles and error
		bars represent the ensemble average and standard
		deviation, and the dotted lines represent the
		theoretical values.}\label{fig:vart}
\end{figure}

The effect of variation in the time step size $\Delta t$ is presented
in Figure~\ref{fig:vart}. Again, we see that the antenna is well
behaved for $\omega_0 \Delta t \lesssim 0.4$. This behavior is expected because we assumed $|\omega_a| \Delta t \ll 1$ in the derivation of the discrete form of the Langevin equation. Note, the very poor
agreement for the FWHM at $\omega_0 \Delta t = \pi / 2$ is due to the
FWHM extending beyond the Nyquist frequency for this case.

The results above demonstrate quantitatively that the antenna is well
behaved for sufficiently small time steps and decorrelation rates. As
discussed in the following section, the decorrelation rates of
interest for plasma turbulence simulations are $\gamma_0 / \omega_0
\lesssim 1$, within the regime of acceptable behavior.  Similarly, 
numerical convergence of simulation results always requires
sufficiently small time steps satisfying $\omega_0 \Delta t \ll
1$. Therefore, our implementation of the oscillating Langevin antenna
is expected to be well behaved for the proposed use of driving plasma
turbulence simulations.

\section{Implementation}\label{sec:imp}

Having defined and characterized the oscillating Langevin antenna in
Section~\ref{sec:ola}, we now describe the use of the antenna for
driving plasma turbulence simulations in the Astrophysical
Gyrokinetics Code, AstroGK.

\subsection{AstroGK}
A detailed description of AstroGK and the results of linear and
nonlinear benchmarks are presented in \cite{Numata:2010}, so we only
provide here a brief overview of the code.

AstroGK is an Eulerian slab code with triply periodic boundary
conditions that solves the electromagnetic gyroaveraged Vlasov-Maxwell
five-dimensional system of equations. It solves the gyrokinetic
equation and gyroaveraged Maxwell's equations for the perturbed
gyroaveraged distribution function, $h_s(x,y,z,\lambda,\epsilon)$, for
each species s, the parallel vector potential $A_z$, the parallel  magnetic
field perturbation $\delta B_z$, and the scalar potential $\phi$
\cite{Frieman:1982,Howes:2006}. The simulation domain is elongated in
the direction of the equilibrium magnetic field.  Velocity space
coordinates are related to the energy, $\epsilon = v^2/2$, and pitch
angle, $\lambda = v_\perp^2 / v^2$. The equilibrium velocity
distribution for all species is treated as Maxwellian, and a realistic
mass ratio, $m_p / m_e = 1836$, is employed for all simulations. The
$x$-$y$~plane is treated pseudospectrally, and an upwinded
finite-differencing approach is employed for the
$z$-direction. Integrals over velocity space are evaluated following
Gaussian quadrature rules. Linear terms are evolved implicitly in
time, while nonlinear terms are evolved explicitly by a third-order
Adams-Bashforth method. Collisions are treated using a fully
conservative, linearized, and gyroaveraged collision operator
\cite{Abel:2008,Barnes:2009}.

\subsection{Antenna Parameter Determination}\label{sec:impl}
Incompressible MHD turbulence is mediated by counterpropagating
Alfv\'{e}n waves since only counterpropagating waves interact
nonlinearly
\cite{Iroshnikov:1963,Kraichnan:1965,Sridhar:1994,Goldreich:1995,Howes:2013a}. If
one adopts the convention that $\omega > 0$, then counterpropagating
waves are described by oppositely signed parallel wave numbers. Also,
for the nonlinearity to be nonzero, the polarization in the
perpendicular plane of the counterpropagating \Alfven waves cannot be
coplanar \cite{Howes:2013a}. Note that these properties make \Alfvenic turbulence
inherently three dimensional
\cite{Howes:2011b,Howes:2012b,Howes:2013a}. We initialize
counterpropagating and perpendicularly polarized Alfv\'{e}n waves in
our simulation by driving the four lowest wavenumber modes in our
domain. If the modes are labelled $(k_x L_\perp, k_y L_\perp, k_z
L_\parallel)$, where the simulation domain is $2\pi (L_\perp, L_\perp,
L_\parallel)$, the driven Fourier modes are  $(1,0,\pm1)$ and
$(0,1,\pm1)$. {Note that the amplitude of each driven mode can be independently specified. By specifying more energy in the field parallel or anti-parallel antenna components, imbalanced turbulence can be generated.} 

An arbitrary number of modes may be driven in AstroGK, however, we
find that driving just these four modes is sufficient to develop a
state of strong \Alfvenic turbulence. Driving only at the smallest wave
numbers in the domain is consistent with the physical model that the
largest scales in our simulation receive energy through nonlinear
interactions between counterpropagating \Alfven waves at scales
slightly larger than the simulation domain.

The condition of critical balance \cite{Goldreich:1995} is used to
determine the appropriate driving amplitude for the antenna to ensure
that a state of strong turbulence is achieved in the simulation.  The
importance of specifying an amplitude sufficient to achieve strong
turbulence cannot be underestimated---many existing plasma turbulence
simulations in the literature do not specify sufficiently large
driving amplitudes (or initial amplitudes for the case of decaying
turbulence simulations), and consequently only a state of weak
turbulence is achieved. Critical balance can be expressed as a balance
between the linear frequency $\omega_l = k_z v_A$ and the nonlinear
frequency, $\omega_l \sim \omega_{nl} = C_2 k_\perp v_\perp$
\cite{Howes:2008b,Howes:2011c}, where $C_2$ is an order unity Kolmogorov constant. The MHD Alfv\'{e}n wave linear
eigenfunction satisfies $v_\perp = \pm
\delta B_\perp / \sqrt{4 \pi n_{0i} m_i}$. At $k_\perp \rho_i \gtrsim
1$, the \Alfven wave transitions into the dispersive kinetic \Alfven
wave (KAW).  More generally, the linear eigenfunction of the
Alfv\'en/KAW satisfies $v_\perp = \overline{\omega}_l \delta B_\perp /
\sqrt{4\pi n_{0i} m_i}$ \cite{Howes:2008b}, where
\begin{equation}
\overline{\omega}_l = \frac{\omega_l}{k_\parallel v_A} = \pm \sqrt{1+\frac{(k_\perp \rho_i)^2}{\beta_i +  2/(1 + T_e/T_i)}}
\end{equation}
is the linear dispersion relation of Alfv\'en/KAWs and $\beta_i =
v_{ti}^2 / v_A^2$. Using the eigenfunction relationship and
\eqref{eq:bperp}, we can express critical balance in terms
of $\tilde{A}_z$, $\omega_l \sim C_2 k_\perp^2 \overline{\omega}_l
\tilde{A}_z/ \sqrt{4 \pi n_{0i} m_i}$, where $\tilde{A}_z$ is the
Fourier coefficient of the parallel vector potential. Therefore,
driving critically balance turbulence at a given scale requires that
the steady-state antenna amplitude satisfy
\begin{equation}\label{eq:crit_amp}
\tilde{A}_z = \frac{\omega_l \sqrt{4 \pi n_{0i} m_i}}{k_\perp^2 C_2 \overline{\omega}_l}.
\end{equation}

As noted in section~\ref{sec:ola}, the long-time response of the
antenna is a Lorentzian. In the presence of a plasma, the long-time
energy response in a particular wavenumber interval of the plasma
remains Lorentzian and takes the form
\begin{equation}
|A_z(t\rightarrow\infty)|^2 = \frac{N A_0^2}{(\overline{\omega}-\overline{\omega}_0)^2 + (\overline{\gamma}_l + \overline{\gamma}_{nl})^2},
\end{equation}
where $N$ is the number of modes driven, overbar indicates
normalization by $k_\parallel v_A$,
$\omega_0$ is the antenna frequency, $\gamma_l$ is the linear damping
rate, and $\gamma_{nl} \simeq \omega_{nl}$ is the nonlinear energy
transfer rate of energy out of the driven wave mode. Since we
typically drive at $k_\perp \rho_i \lesssim 1$, the linear damping
rate will be much smaller than the nonlinear energy transfer rate:
$\gamma_l \ll \omega_{nl}$. Also, $\omega_l \sim \omega_{nl}$ in
critical balance. Due to the shifting frequency of the antenna,
$\overline{\omega} - \overline{\omega}_0 \sim [0,1]
\overline{\omega}_{nl} \sim [0,1] \overline{\omega}_{l}$. To account
for this, we introduce a parameter $\delta \in [1,2]$ so that the
denominator becomes $\overline{\omega}_l^2 \delta$. Therefore,
\begin{equation}\label{eq:lor_final}
|A_z(t\rightarrow\infty)| = \frac{\sqrt{N} A_0}{\overline{\omega}_l \sqrt{\delta} }.
\end{equation}

Equating \eqref{eq:crit_amp} and \eqref{eq:lor_final}, we
arrive at the final form for the necessary  driving amplitude to
achieve a state of critically balanced, strong turbulence
\begin{equation}\label{eq:final_A0}
A_0 = \frac{\overline{\omega}_l B_0 k_\parallel\sqrt{\delta} }{C_2 k_\perp^2 \sqrt{N}}.
\end{equation}
We typically take $\delta = 2$ and $C_2 = 1$ to evaluate \eqref{eq:final_A0}. 

Finally, we need to specify the driving frequency, $\omega_0$, and
decorrelation rate, $\gamma_0$ of the antenna. Since the energy entering our
simulation is meant to mimic turbulence cascaded from larger scales,
the frequency of the input energy should be lower than the linear
frequency of the driven mode. Thus, we choose a driving frequency
slightly below the characteristic linear frequency of the plasma at
the driving wavelength, $\omega_l = k_\parallel v_A
\overline{\omega}_l$. We typically take $\omega_0 = 0.9 \omega_l$. We
also want a decorrelation rate of order the linear frequency but below the antenna frequency since the nonlinear cascade rate at a given scale is expected to be equal to or less than the linear frequency at that scale. Therefore, $\gamma_0 \le \omega_0$ based upon physical expectations, and we typically choose $\gamma_0 = -0.8 \omega_l$.

Although we have focused here on a discussion of driving strong
turbulence in our domain, by reducing the driving amplitude $A_0$ to a
value less than that given by \eqref{eq:final_A0}, we can
intentionally drive weak turbulence. \citet{Perez:2008} noted that an
elongated simulation domain is necessary for the optimal study of weak
turbulence, which allows them to use reduced MHD simulations. Reduced
MHD is indeed a limit of gyrokinetics \cite{Schekochihin:2009}. As part of its ordering,
gyrokinetics necessitates the use of an elongated simulation domain,
which implies that the assumed ordering underlying our simulations can
also be used for studying weak turbulence.  

\subsection{Saturation of Strong Turbulence in AstroGK Simulations}\label{sec:agk_sim}

\begin{figure}[t]
		\includegraphics[width=\linewidth]{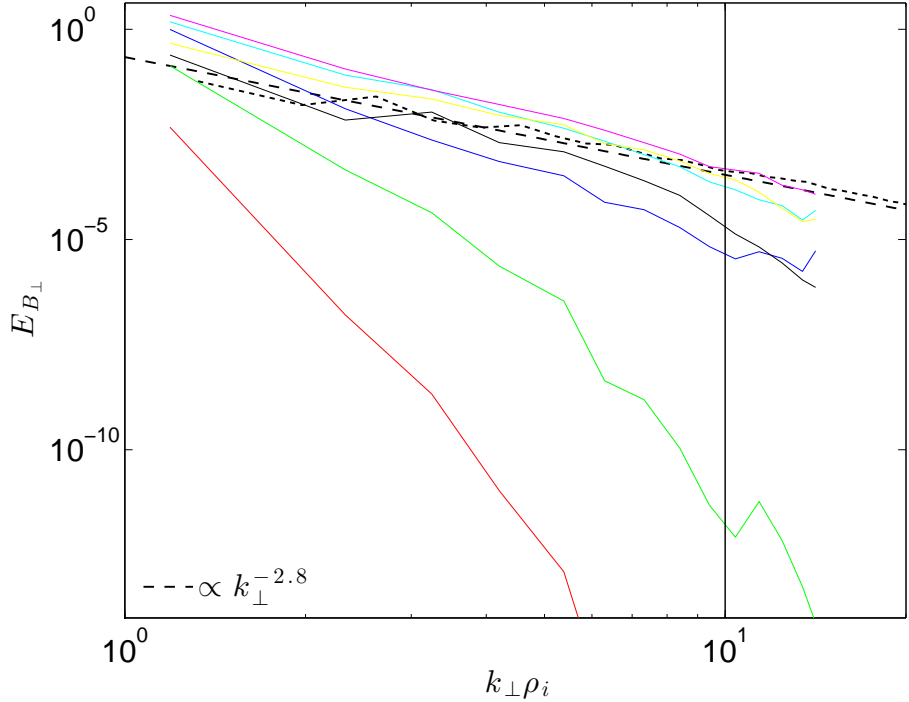}
		\caption{Evolution of the perpendicular magnetic
		energy spectrum for a $\beta_i = 1$ fully nonlinear
		AstroGK simulation driven as outlined in section
		\ref{sec:impl}. The solid lines represent the energy
		at $t \simeq 0.1, 0.2, 0.4, 0.8, 1.6, 3.2,$ and $6.4
		\tau_{D}$ (red, green, blue, cyan, magenta,
		yellow, and black respectively) and the dotted line is
		from a larger, higher resolution simulation with
		similar plasma parameters. The vertical line at
		$k_\perp \rho_i = 10$ represents the edge of the
		simulation domain.}\label{fig:kspec_evol}
\end{figure}

Driving in the manner described above generates well-developed
turbulence across the full simulation domain within approximately one
domain-scale turnaround time: $\tau_{D} = 2\pi /
\omega_{D}$. Since we choose to drive in critical balance,
$\tau_{D} = \tau_{Alfven} = 2 \pi / v_A k_{z D}$, where $k_{z D} = 2
\pi / L_z$. To demonstrate this, we plot in
Figure~\ref{fig:kspec_evol} the evolution of the perpendicular
magnetic energy spectrum, $E_{B_\perp} = \delta B_\perp^2 / k_\perp$,
spanning $0.1$ to $6.4 \tau_{D}$ for a simulation of strong \Alfvenic
turbulence in a $\beta_i = 1$ plasma using AstroGK driven by the
oscillating Langevin antenna as described above---full details of the
simulation are provided in \cite{TenBarge:2012a}. The specific
parameters of the antenna for this simulation are: $\omega_0 = 1.14
k_{z D} v_A \simeq \omega_l$, $\gamma_0 = -0.9 k_{z D} v_A$, $A_0 = \rho_i B_0 / \xi$,
and with four driven wavenumber modes $(k_{x}, {k}_{y}, {k}_{z} / \xi)
\rho_i = (1,0,\pm1)$ and $(0,1,\pm1)$, where the spatial extent of the
domain is $(L_{x}, L_{y}, L_{z}) = 2 \pi \rho_i (1, 1, 1/\xi)$ and
$\xi$ is an elongation factor. The time step at the
beginning of the simulation was $\Delta t = 5 \times 10^{-3} / k_{z D}
v_A$ but decreased to $\Delta t \simeq 2 \times 10^{-5} / k_{z D} v_A$
during the course of the simulation to ensure the
Courant-Friedrichs-Lewy (CFL) condition is satisfied at the smallest
spatial scales. The only energy input into the system is via the
oscillating Langevin antenna. Within one turnaround time, the spectrum
develops an approximate $k_\perp^{-2.8}$ scaling, consistent with
solar wind observations
\cite{Kiyani:2009,Alexandrova:2009,Sahraoui:2010b}.

In addition, we plot the magnetic energy spectrum from a
previously published AstroGK simulation over the entire dissipation
range from the ion to the electron Larmor radius scale
\cite{Howes:2011b}. This simulation employs the same  plasma parameters
as employed above, but covers a much larger spatial dynamic range.
The antenna and plasma parameters of the larger simulation are
identical to the smaller simulation with the following exceptions: six
modes were driven, $(k_{x}, {k}_{y}, {k}_{z} / \xi)
\rho_i = (\pm1,0,\pm1)$ and $(0,1,\pm1)$, and the ending time step was
$\Delta t \simeq 2.4 \times 10^{-6} /k_{z D} v_A$.  We see that the 
results of this larger  simulation (dotted) agree well with the 
results of the smaller simulation in Figure~\ref{fig:kspec_evol}.

\subsection{Comparing Turbulent Driving to Fluctuations in Turbulence Simulations}\label{sec:agk_sim}

\begin{figure}[t]
		\includegraphics[width=\linewidth]{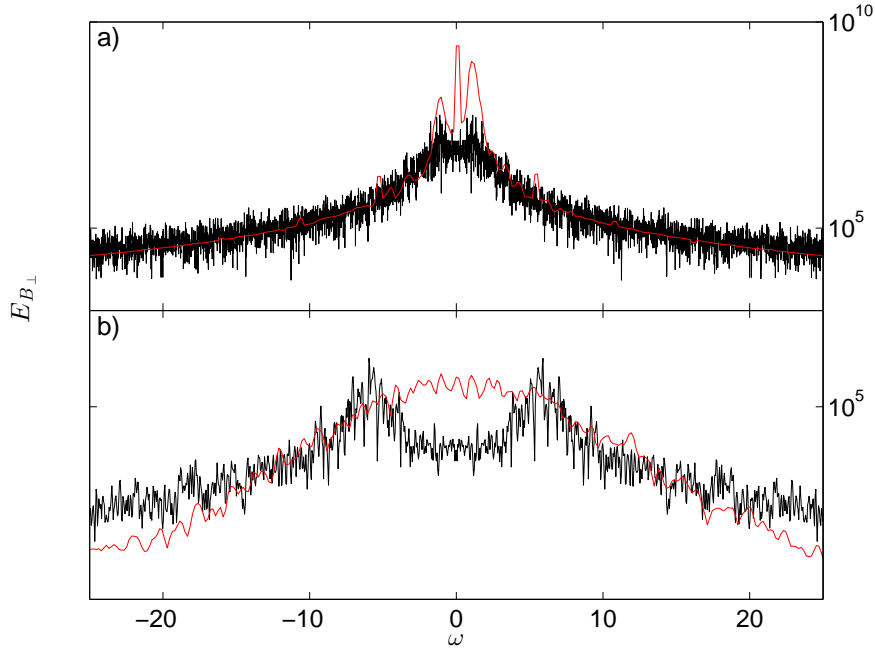}
		\caption{Plasma response to the oscillating Langevin
		antenna. In both panels, the plasma response is
		plotted in red and the antenna is black. Panel a) is
		the response for the driven mode with $k_\perp \rho_i
		= 1$ and $k_z \rho_i / \xi = 1$. The antenna
		parameters are the same as used in the simulation,
		$\omega_0 = \pm 1.14$ and $\gamma_0 = -0.9$. Panel b)
		is the response in the middle of the domain with
		$k_\perp \rho_i = 5$ and $k_z \rho_i / \xi = 1$. The
		antenna parameters are $\omega_0 = \pm 6$ and
		$\gamma_0 = -\omega_0/2\pi$.}\label{fig:ant_resp}
\end{figure}

In Figure~\ref{fig:ant_resp}, we examine the plasma response to the
oscillating Langevin antenna for the same small simulation discussed
above. In both panels, the red line is the plasma response, and the
black line is the sum of antennae with positive and negative
frequencies to represent waves driven up and down the magnetic
field. Both quantities have applied to them a boxcar average with a
width of 2. In panel a) of the figure is the driven mode of the plasma
with $k_\perp \rho_i = 1$ and $k_z \rho_i / \xi = 1$, and the antenna
has the same parameters as those used in the simulation, $\omega_0 =
\pm 1.14 k_{zD} v_A$ and $\gamma_0 = -0.9 k_{zD} v_A$. The linear frequency of this plasma
mode is $\omega_l = 1.14 k_{zD} v_A$, which corresponds to the driving
frequency. Clearly, the plasma response of this mode is dominated by
the driving. It can also be seen that more energy is driven into the
positive frequency mode, which implies somewhat imbalanced
driving. The accumulation of energy in the $\omega = 0$ mode is
expected because this mode is responsible for nonlinear scattering in
three-wave interactions of turbulence and is self-consistently
generated
\cite{Montgomery:1981,Sridhar:1994,Galtier:2003,TenBarge:2012a,Howes:2013a}.

In panel b) of Figure \ref{fig:ant_resp} is the plasma response in the
middle of the perpendicular domain, $k_\perp \rho_i = 5$ and and $k_z
\rho_i / \xi = 1$. The simulation is not driven at this wavenumber, but it is valuable to compare the natural plasma response to that of the oscillating Langevin antenna to determine if the antenna does indeed resemble turbulence at a given scale. The linear frequency associated with this mode is
$\omega \simeq 6$, so we model the plasma response with an antenna
having a frequency $\omega_0 = 6$. We choose an antenna decorrelation
rate $\gamma_0 = -\omega_0 / 2\pi$ to fit the response. Although the
antenna fit to the plasma appears poor, the subjective appearance is
deceptive. The discussion in \S \ref{sec:drive} suggests the plasma
will respond poorly to energy input above the linear frequency of a
given mode. Therefore, the apparent excess energy in the tails of the
antenna will couple poorly to the plasma. Further, the valley between
the antenna peaks will be self-consistently populated by low frequency
modes generated in critically balanced turbulence \cite{Howes:2013a}.


\section{Comparison to Other Driving Methods}\label{sec:compare}

We have described one method by which turbulence simulations can be
driven; however, many methods are used throughout the literature to
initialize or drive turbulence. We now briefly discuss the two most common approaches to generating plasma turbulence in numerical simulations. 


\subsection{Decaying}

One common method to inject energy into turbulence simulations is to
initialize a set of modes at the beginning of the simulation and
observe how they decay and energy is cascaded to smaller scales over
time. Decaying simulations of turbulence can be initialized in a
variety of ways to facilitate the study of various phenomena, for
example: generalized Orszag-Tang vortices to simulate magnetic
reconnection driven turbulence in electron MHD (EMHD)
\cite{Biskamp:1999} or hybrid simulations \cite{Parashar:2009}, a
spectrum of energy across a band of Fourier modes with varying angular
distributions in EMHD \cite{Dastgeer:2000}, exact plasma eigenmodes
with equal energy in particle-in-cell simulations
\cite{Saito:2008,Svidzinski:2009}, energy in only the largest Fourier
mode in gyrokinetic simulations \cite{Tatsuno:2010}, equal energy
across a collection of large scale Fourier modes in Landau fluid
simulations \cite{Hunana:2011}. 

Decaying turbulence simulations have the advantage of specifying an
exact initial condition and energy state, which provide precise
control of the simulation. This is ideal when the goal of the
simulation is to either study the evolution of exact plasma eigenmodes
or a specific physical configuration (e.g., Orszag-Tang vortices);
however, it is unlikely to represent many physical systems since a
collection of isolated exact eigenmodes or an ideal spatial
configuration are rare events. Therefore, decaying simulations tend to
represent highly idealized physical processes. Also, decaying simulations are incapable of achieving a steady-state by definition.

\subsection{Direct Injection of Noise}

Direct injection of noise has been used to drive many turbulence
simulations.  The simulations employ a variety of methods for
injecting energy; however, they all do so by setting the Fourier
coefficients of the velocity and/or magnetic field to random numbers
at either each time step or over finitely correlated times for a band
of wave vectors with appropriate normalizations and constraints, e.g.,
$\nabla \cdot \bB = 0$, applied to the Fourier coefficients prior to
advancing the fields. The Fourier coefficients are typically updated by
randomizing only the phase with fixed amplitude integrated over all
driven wavenumbers \cite{Cho:2000,Perez:2008,Beresnyak:2009,Cho:2009},
randomizing the phase and amplitude within a Gaussian envelope
\cite{Maron:2001}, or choosing random phase, amplitude, and driving
wavenumber from a band of wavenumbers
\cite{Haugen:2003}. {\citet{Muller:2005} drive a simulation by initializing and "freezing" modes in a given band of wavenumbers, which is similar to the method used in decaying turbulence simulations except that the initial condition is held fixed.}

As demonstrated by its ubiquity, this method of driving is capable of
generating strong turbulence; however, the physical motivation for the
various driving methods outlined above is unlikely to be
representative of large scale energy being cascaded into the
simulation domain. Although useful from an analysis standpoint,
injecting energy with fixed amplitude and constant or varying phase
with respect to time is not well motivated physically. Randomizing the
phase and amplitude from a range of Gaussian distributed amplitudes is
likely more realistic, but still leaves unspecified the frequency spectrum of the
energy input into the system.

\begin{figure}[t]
		\includegraphics[width=\linewidth]{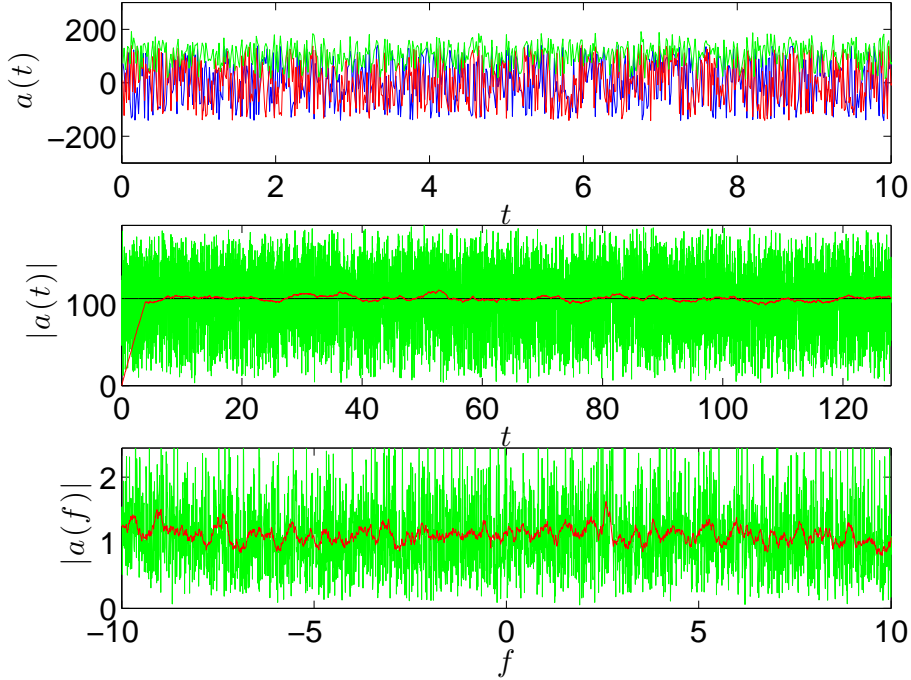}
		\caption{Temporal evolution of a single,
		delta-correlated white noise antenna with $A_0 =
		100$. Top: Real (blue), imaginary (red), and the
		magnitude (green) of the complex vector potential over
		10 periods. Middle: Magnitude (green), boxcar average
		of the magnitude (red), and overall average of the
		magnitude (black) of $a$ over 128 periods. Bottom: The
		frequency spectrum of the antenna (green) and its
		boxcar average (red).}\label{fig:white}
\end{figure}

Many simulations employ finite time correlated driving wherein the
Fourier coefficients are updated at some fraction of the system
crossing time. For this type of driving, the frequency of the energy
input is difficult to determine without greater detail than is
typically provided. However, simulations that employ a
delta-correlated driving correspond to directly injecting white noise
in frequency. White noise driving can be described by an equation
similar to equation~\eqref{eq:lang_cont} in the absence of oscillations
and damping,
\begin{equation}
\frac{da}{dt} = \sqrt{2} A_0 u_n.
\end{equation}
The evolution of this type of antenna is depicted in Figure
\ref{fig:white}. The ensemble average of 64 similar runs for driving
of this type is $<|a|> = 108.1649 \pm 40.28515$. As expected, the
energy injected by white noise is evenly distributed across all
frequencies. This will result in a magnetic energy spectrum $E(\omega)
\propto \delta B^2 / \omega \propto \omega^{-1}$.



White noise driving might be motivated by an attempt to mimic the
energy containing region of turbulence since the energy containing
region is expected to be inhomogeneous and have an energy spectrum
that scales as $k^{-1}$. Although the spectral exponent is $-1$, this
is due to a constant energy distribution in space. The temporal
distribution of energy in this region is not known but is unlikely to
be constant. Note that solar wind observations of this region depict a
frequency spectrum with spectral exponent $-1$; however, this is a
measurement of spatial plasma fluctuations and not temporal plasma
oscillations due to the high Mach number flows and single point
measurements of the solar wind, e.g., see \cite{Bruno:2005}.

Most of the methods outlined in this section reach a steady-state
within a few turnaround times, so they are efficient at developing
strong turbulence. Despite the rapidity with which they develop strong
turbulence, the methods tend to pollute wavenumbers approximately a
factor of two or more beyond their driven range with driving
effects. For instance, \citet{Perez:2008} drive in the perpendicular
direction with $1 \le k_\perp \le 2$, but the driving can be seen to
affect perpendicular wavenumbers up to $k_\perp \lesssim 4$.

\section{Summary}\label{sec:summary}

We have presented a novel method for driving plasma turbulence simulations via an oscillating Langevin antenna that is evolved via the Langevin equation. This method of driving plasma turbulence was motivated by a desire to perform turbulence simulations in which the largest scale of the simulation is smaller than the scale at which energy is physically injected into the plasma environment that is being modeled. In other words, the outer scale of the inertial range of the physical turbulent system is larger than the simulation domain scale. An overview of the properties of plasma turbulence relevant to driving in a physical manner were explored in \S \ref{sec:turb}. Also, the response of the plasma to simple sinusoidal driving was considered in \S \ref{sec:drive}, where we found that the driving frequency of the antenna must be below the characteristic linear frequency of the plasma to avoid an impedance mismatch. 

Driving turbulence
via the oscillating Langevin antenna requires counterpropagating Alfven waves, in which each driven plane Alfven wave is completely characterized by four parameters: the wavevector $\V{k}$, amplitude $A_0$, frequency $\omega_0$, and decorrelation rate $\gamma_0$. We have shown that the oscillating Langevin antenna represents an efficient and flexible method to drive
\Alfvenic turbulence simulations that mimics cascaded turbulent energy
entering a numerical simulation domain from scales larger than those included in the
simulation. By varying the antenna amplitude, we are able to drive
strong or weak turbulence into the simulation domain, and the turbulence can be
balanced or imbalanced.

The domain of validity of the antenna was explored in \S
\ref{sec:ola}, where we found that the antenna is well behaved for
de-correlation rates $| \gamma_0| \lesssim 4 \omega_0$ and time steps
$\Delta t \lesssim 0.4 / \omega_0$. For all cases of interest, the nonlinear cascade rate is less than or equal to the linear frequency, so $|\gamma_0| \lesssim 4
\omega_0$ is always satisfied. Also, $\Delta t \ll 1 / \omega_l$ to
ensure numerical accuracy of any simulation, where $\omega_l$ is the linear response of the plasma. Therefore, the antenna is expected to be well behaved in any reasonable turbulence simulation.

In \S \ref{sec:compare}, we examined two of the most common methods, decaying and injection of noise, of generating turbulence in numerical simulations and compared them to the oscillating Langevin antenna. We found that both of these common methods have certain inherent advantages but that neither is well motivated physically nor does either method represent a realistic frequency response of energy entering the simulation domain.

Although our presentation of the oscillating Langevin antenna has focused on its usage to drive \Alfvenic turbulence in the Astrophysical Gyrokinetics Code, the prescription for the antenna given in \S \ref{sec:ola} is sufficiently generic to permit implementation of similar schemes in other numerical simulations of plasma turbulence.

\section*{Acknowledgements}

 Support was provided by NSF CAREER Award AGS-1054061 and NSF grant
PHY-10033446.


\begin{thebibliography}{83}
\providecommand{\natexlab}[1]{#1}
\providecommand{\url}[1]{\texttt{#1}}
\providecommand{\urlprefix}{URL }
\expandafter\ifx\csname urlstyle\endcsname\relax
  \providecommand{\doi}[1]{doi:\discretionary{}{}{}#1}\else
  \providecommand{\doi}[1]{doi:\discretionary{}{}{}\begingroup
  \urlstyle{rm}\url{#1}\endgroup}\fi
\providecommand{\bibinfo}[2]{#2}

\bibitem[{{Belcher} and {Davis}(1971)}]{Belcher:1971}
\bibinfo{author}{J.~W. {Belcher}}, \bibinfo{author}{L.~{Davis}},
  \bibinfo{title}{{Large-Amplitude Alfv{\'e}n Waves in the Interplanetary
  Medium, 2}}, \bibinfo{journal}{J.~Geophys.~Res.} \bibinfo{volume}{76}
  (\bibinfo{year}{1971}) \bibinfo{pages}{3534--3563}.

\bibitem[{{Tu} and {Marsch}(1995)}]{Tu:1995}
\bibinfo{author}{C.-Y. {Tu}}, \bibinfo{author}{E.~{Marsch}},
  \bibinfo{title}{{MHD structures, waves and turbulence in the solar wind:
  Observations and theories}}, \bibinfo{journal}{Space Science Reviews}
  \bibinfo{volume}{73} (\bibinfo{year}{1995}) \bibinfo{pages}{1--2}.

\bibitem[{{Bruno} and {Carbone}(2005)}]{Bruno:2005}
\bibinfo{author}{R.~{Bruno}}, \bibinfo{author}{V.~{Carbone}},
  \bibinfo{title}{{The Solar Wind as a Turbulence Laboratory}},
  \bibinfo{journal}{Living Reviews in Solar Physics}
  \bibinfo{volume}{2}~(\bibinfo{number}{4}).

\bibitem[{Iroshnikov(1963)}]{Iroshnikov:1963}
\bibinfo{author}{R.~S. Iroshnikov}, \bibinfo{title}{The turbulence of a
  conducting fluid in a strong magnetic field}, \bibinfo{journal}{Astron. Zh.}
  \bibinfo{volume}{40} (\bibinfo{year}{1963}) \bibinfo{pages}{742},
  \bibinfo{note}{{English} Translation: Sov. Astron., 7 566 (1964)}.

\bibitem[{Kraichnan(1965)}]{Kraichnan:1965}
\bibinfo{author}{R.~H. Kraichnan}, \bibinfo{title}{Inertial Range Spectrum of
  Hyromagnetic Turbulence}, \bibinfo{journal}{Phys.~Fluids} \bibinfo{volume}{8}
  (\bibinfo{year}{1965}) \bibinfo{pages}{1385--1387}.

\bibitem[{Shebalin et~al.(1983)Shebalin, Matthaeus, and
  Montgomery}]{Shebalin:1983}
\bibinfo{author}{J.~V. Shebalin}, \bibinfo{author}{W.~H. Matthaeus},
  \bibinfo{author}{D.~Montgomery}, \bibinfo{title}{Anisotropy in MHD turbulence
  due to a mean magnetic field}, \bibinfo{journal}{J.~Plasma Phys.}
  \bibinfo{volume}{29} (\bibinfo{year}{1983}) \bibinfo{pages}{525--547}.

\bibitem[{Sridhar and Goldreich(1994)}]{Sridhar:1994}
\bibinfo{author}{S.~Sridhar}, \bibinfo{author}{P.~Goldreich},
  \bibinfo{title}{{Toward a Thoery of Interstellar Turbulence I. Weak
  Alfv\'enic Turbulence}}, \bibinfo{journal}{Astrophys.~J.}
  \bibinfo{volume}{433} (\bibinfo{year}{1994}) \bibinfo{pages}{612--621}.

\bibitem[{Goldreich and Sridhar(1995)}]{Goldreich:1995}
\bibinfo{author}{P.~Goldreich}, \bibinfo{author}{S.~Sridhar},
  \bibinfo{title}{{Toward a Thoery of Interstellar Turbulence II. Strong
  Alfv\'enic Turbulence}}, \bibinfo{journal}{Astrophys.~J.}
  \bibinfo{volume}{438} (\bibinfo{year}{1995}) \bibinfo{pages}{763--775}.

\bibitem[{{Boldyrev}(2006)}]{Boldyrev:2006}
\bibinfo{author}{S.~{Boldyrev}}, \bibinfo{title}{{Spectrum of
  Magnetohydrodynamic Turbulence}}, \bibinfo{journal}{Phys.~Rev.~Lett.}
  \bibinfo{volume}{96}~(\bibinfo{number}{11}) (\bibinfo{year}{2006})
  \bibinfo{pages}{115002--+},
  \doi{\bibinfo{doi}{10.1103/PhysRevLett.96.115002}}.

\bibitem[{{Howes} and {Nielson}(2013)}]{Howes:2013a}
\bibinfo{author}{G.~G. {Howes}}, \bibinfo{author}{K.~D. {Nielson}},
  \bibinfo{title}{{Alfven Wave Collisions, The Fundamental Building Block of
  Plasma Turbulence I: Asymptotic Solution}}, \bibinfo{journal}{Phys.~Plasmas}
  \bibinfo{volume}{20} (\bibinfo{year}{2013}) \bibinfo{pages}{072302}.

\bibitem[{{Taylor}(1938)}]{Taylor:1938}
\bibinfo{author}{G.~I. {Taylor}}, \bibinfo{title}{{The Spectrum of
  Turbulence}}, \bibinfo{journal}{{Proc. Roy. Soc. A}} \bibinfo{volume}{164}
  (\bibinfo{year}{1938}) \bibinfo{pages}{476--490}.

\bibitem[{{Howes} et~al.(2012){Howes}, {Drake}, {Nielson}, {Carter},
  {Kletzing}, and {Skiff}}]{Howes:2012b}
\bibinfo{author}{G.~G. {Howes}}, \bibinfo{author}{D.~J. {Drake}},
  \bibinfo{author}{K.~D. {Nielson}}, \bibinfo{author}{T.~A. {Carter}},
  \bibinfo{author}{C.~A. {Kletzing}}, \bibinfo{author}{F.~{Skiff}},
  \bibinfo{title}{{Toward Astrophysical Turbulence in the Laboratory}},
  \bibinfo{journal}{Phys.~Rev.~Lett.}
  \bibinfo{volume}{109}~(\bibinfo{number}{25}) \bibinfo{eid}{255001},
  \doi{\bibinfo{doi}{10.1103/PhysRevLett.109.255001}}.

\bibitem[{{Gekelman}(1999)}]{Gekelman:1999}
\bibinfo{author}{W.~{Gekelman}}, \bibinfo{title}{{Review of laboratory
  experiments on Alfv{\'e}n waves and their relationship to space
  observations}}, \bibinfo{journal}{J.~Geophys.~Res.} \bibinfo{volume}{104}
  (\bibinfo{year}{1999}) \bibinfo{pages}{14417--14436},
  \doi{\bibinfo{doi}{10.1029/98JA00161}}.

\bibitem[{{Howes} et~al.(2008{\natexlab{a}}){Howes}, {Cowley}, {Dorland},
  {Hammett}, {Quataert}, and {Schekochihin}}]{Howes:2008b}
\bibinfo{author}{G.~G. {Howes}}, \bibinfo{author}{S.~C. {Cowley}},
  \bibinfo{author}{W.~{Dorland}}, \bibinfo{author}{G.~W. {Hammett}},
  \bibinfo{author}{E.~{Quataert}}, \bibinfo{author}{A.~A. {Schekochihin}},
  \bibinfo{title}{{A model of turbulence in magnetized plasmas: Implications
  for the dissipation range in the solar wind}},
  \bibinfo{journal}{J.~Geophys.~Res.} \bibinfo{volume}{113}
  (\bibinfo{year}{2008}{\natexlab{a}}) \bibinfo{pages}{A05103},
  \doi{\bibinfo{doi}{10.1029/2007JA012665}}.

\bibitem[{{Howes}(2008)}]{Howes:2008c}
\bibinfo{author}{G.~G. {Howes}}, \bibinfo{title}{{Inertial range turbulence in
  kinetic plasmas}}, \bibinfo{journal}{Phys.~Plasmas}
  \bibinfo{volume}{15}~(\bibinfo{number}{5}) (\bibinfo{year}{2008})
  \bibinfo{pages}{055904}, \doi{\bibinfo{doi}{10.1063/1.2889005}}.

\bibitem[{{Schekochihin} et~al.(2009){Schekochihin}, {Cowley}, {Dorland},
  {Hammett}, {Howes}, {Quataert}, and {Tatsuno}}]{Schekochihin:2009}
\bibinfo{author}{A.~A. {Schekochihin}}, \bibinfo{author}{S.~C. {Cowley}},
  \bibinfo{author}{W.~{Dorland}}, \bibinfo{author}{G.~W. {Hammett}},
  \bibinfo{author}{G.~G. {Howes}}, \bibinfo{author}{E.~{Quataert}},
  \bibinfo{author}{T.~{Tatsuno}}, \bibinfo{title}{{Astrophysical Gyrokinetics:
  Kinetic and Fluid Turbulent Cascades in Magnetized Weakly Collisional
  Plasmas}}, \bibinfo{journal}{Astrophys.~J.~Supp.} \bibinfo{volume}{182}
  (\bibinfo{year}{2009}) \bibinfo{pages}{310--377},
  \doi{\bibinfo{doi}{10.1088/0067-0049/182/1/310}}.

\bibitem[{{Parashar} et~al.(2009){Parashar}, {Shay}, {Cassak}, and
  {Matthaeus}}]{Parashar:2009}
\bibinfo{author}{T.~N. {Parashar}}, \bibinfo{author}{M.~A. {Shay}},
  \bibinfo{author}{P.~A. {Cassak}}, \bibinfo{author}{W.~H. {Matthaeus}},
  \bibinfo{title}{{Kinetic dissipation and anisotropic heating in a turbulent
  collisionless plasma}}, \bibinfo{journal}{Phys.~Plasmas}
  \bibinfo{volume}{16}~(\bibinfo{number}{3}) (\bibinfo{year}{2009})
  \bibinfo{pages}{032310--+}, \doi{\bibinfo{doi}{10.1063/1.3094062}}.

\bibitem[{{Osman} et~al.(2011){Osman}, {Matthaeus}, {Greco}, and
  {Servidio}}]{Osman:2011}
\bibinfo{author}{K.~T. {Osman}}, \bibinfo{author}{W.~H. {Matthaeus}},
  \bibinfo{author}{A.~{Greco}}, \bibinfo{author}{S.~{Servidio}},
  \bibinfo{title}{{Evidence for Inhomogeneous Heating in the Solar Wind}},
  \bibinfo{journal}{Astrophys.~J.~Lett.} \bibinfo{volume}{727}
  \bibinfo{eid}{L11}, \doi{\bibinfo{doi}{10.1088/2041-8205/727/1/L11}}.

\bibitem[{{Servidio} et~al.(2012){Servidio}, {Valentini}, {Califano}, and
  {Veltri}}]{Servidio:2012}
\bibinfo{author}{S.~{Servidio}}, \bibinfo{author}{F.~{Valentini}},
  \bibinfo{author}{F.~{Califano}}, \bibinfo{author}{P.~{Veltri}},
  \bibinfo{title}{{Local Kinetic Effects in Two-Dimensional Plasma
  Turbulence}}, \bibinfo{journal}{Phys.~Rev.~Lett.}
  \bibinfo{volume}{108}~(\bibinfo{number}{4}) \bibinfo{eid}{045001},
  \doi{\bibinfo{doi}{10.1103/PhysRevLett.108.045001}}.

\bibitem[{{TenBarge} and {Howes}(2013)}]{TenBarge:2013a}
\bibinfo{author}{J.~M. {TenBarge}}, \bibinfo{author}{G.~G. {Howes}},
  \bibinfo{title}{{Current Sheets and Collisionless Dissipation in Kinetic
  Plasma Turbulence}}, \bibinfo{journal}{Astrophys.~J.~Lett.}
  \bibinfo{volume}{771} (\bibinfo{year}{2013}) \bibinfo{pages}{L27}.

\bibitem[{{Karimabadi} et~al.(2013){Karimabadi}, {Roytershteyn}, {Wan},
  {Matthaeus}, {Daughton}, {Wu}, {Shay}, {Loring}, {Borovsky}, {Leonardis},
  {Chapman}, and {Nakamura}}]{Karimabadi:2013}
\bibinfo{author}{H.~{Karimabadi}}, \bibinfo{author}{V.~{Roytershteyn}},
  \bibinfo{author}{M.~{Wan}}, \bibinfo{author}{W.~H. {Matthaeus}},
  \bibinfo{author}{W.~{Daughton}}, \bibinfo{author}{P.~{Wu}},
  \bibinfo{author}{M.~{Shay}}, \bibinfo{author}{B.~{Loring}},
  \bibinfo{author}{J.~{Borovsky}}, \bibinfo{author}{E.~{Leonardis}},
  \bibinfo{author}{S.~C. {Chapman}}, \bibinfo{author}{T.~K.~M. {Nakamura}},
  \bibinfo{title}{{Coherent structures, intermittent turbulence, and
  dissipation in high-temperature plasmas}}, \bibinfo{journal}{Phys.~Plasmas}
  \bibinfo{volume}{20}~(\bibinfo{number}{1}) (\bibinfo{year}{2013})
  \bibinfo{pages}{012303}, \doi{\bibinfo{doi}{10.1063/1.4773205}}.

\bibitem[{{Coleman}(1968)}]{Coleman:1968}
\bibinfo{author}{P.~J. {Coleman}, Jr.}, \bibinfo{title}{{Turbulence, Viscosity,
  and Dissipation in the Solar-Wind Plasma}}, \bibinfo{journal}{Astrophys.~J.}
  \bibinfo{volume}{153} (\bibinfo{year}{1968}) \bibinfo{pages}{371--388}.

\bibitem[{Leamon et~al.(1998)Leamon, Smith, Ness, Matthaeus, and
  Wong}]{Leamon:1998a}
\bibinfo{author}{R.~J. Leamon}, \bibinfo{author}{C.~W. Smith},
  \bibinfo{author}{N.~F. Ness}, \bibinfo{author}{W.~H. Matthaeus},
  \bibinfo{author}{H.~K. Wong}, \bibinfo{title}{Observational Constraints on
  the Dynamics of the Interplanetary Magnetic Field Dissipation Range},
  \bibinfo{journal}{J.~Geophys.~Res.} \bibinfo{volume}{103}
  (\bibinfo{year}{1998}) \bibinfo{pages}{4775--4787}.

\bibitem[{{Sahraoui} et~al.(2009){Sahraoui}, {Goldstein}, {Robert}, and
  {Khotyaintsev}}]{Sahraoui:2009}
\bibinfo{author}{F.~{Sahraoui}}, \bibinfo{author}{M.~L. {Goldstein}},
  \bibinfo{author}{P.~{Robert}}, \bibinfo{author}{Y.~V. {Khotyaintsev}},
  \bibinfo{title}{{Evidence of a Cascade and Dissipation of Solar-Wind
  Turbulence at the Electron Gyroscale}}, \bibinfo{journal}{Physical Review
  Letters} \bibinfo{volume}{102}~(\bibinfo{number}{23}) (\bibinfo{year}{2009})
  \bibinfo{pages}{231102--+},
  \doi{\bibinfo{doi}{10.1103/PhysRevLett.102.231102}}.

\bibitem[{{Kiyani} et~al.(2009){Kiyani}, {Chapman}, {Khotyaintsev}, {Dunlop},
  and {Sahraoui}}]{Kiyani:2009}
\bibinfo{author}{K.~H. {Kiyani}}, \bibinfo{author}{S.~C. {Chapman}},
  \bibinfo{author}{Y.~V. {Khotyaintsev}}, \bibinfo{author}{M.~W. {Dunlop}},
  \bibinfo{author}{F.~{Sahraoui}}, \bibinfo{title}{{Global scale-invariant
  dissipation in collisionless plasma turbulence}}, \bibinfo{journal}{Physical
  Review Letters} \bibinfo{volume}{103} (\bibinfo{year}{2009})
  \bibinfo{pages}{075006}.

\bibitem[{{Alexandrova} et~al.(2009){Alexandrova}, {Saur}, {Lacombe},
  {Mangeney}, {Mitchell}, {Schwartz}, and {Robert}}]{Alexandrova:2009}
\bibinfo{author}{O.~{Alexandrova}}, \bibinfo{author}{J.~{Saur}},
  \bibinfo{author}{C.~{Lacombe}}, \bibinfo{author}{A.~{Mangeney}},
  \bibinfo{author}{J.~{Mitchell}}, \bibinfo{author}{S.~J. {Schwartz}},
  \bibinfo{author}{P.~{Robert}}, \bibinfo{title}{{Universality of Solar-Wind
  Turbulent Spectrum from MHD to Electron Scales}},
  \bibinfo{journal}{Phys.~Rev.~Lett.}
  \bibinfo{volume}{103}~(\bibinfo{number}{16}) (\bibinfo{year}{2009})
  \bibinfo{pages}{165003--+},
  \doi{\bibinfo{doi}{10.1103/PhysRevLett.103.165003}}.

\bibitem[{{Chen} et~al.(2010){Chen}, {Horbury}, {Schekochihin}, {Wicks},
  {Alexandrova}, and {Mitchell}}]{Chen:2010}
\bibinfo{author}{C.~H.~K. {Chen}}, \bibinfo{author}{T.~S. {Horbury}},
  \bibinfo{author}{A.~A. {Schekochihin}}, \bibinfo{author}{R.~T. {Wicks}},
  \bibinfo{author}{O.~{Alexandrova}}, \bibinfo{author}{J.~{Mitchell}},
  \bibinfo{title}{{Anisotropy of Solar Wind Turbulence between Ion and Electron
  Scales}}, \bibinfo{journal}{Phys.~Rev.~Lett.}
  \bibinfo{volume}{104}~(\bibinfo{number}{25}) (\bibinfo{year}{2010})
  \bibinfo{pages}{255002--+},
  \doi{\bibinfo{doi}{10.1103/PhysRevLett.104.255002}}.

\bibitem[{{Sahraoui} et~al.(2010){Sahraoui}, {Goldstein}, {Belmont}, {Canu},
  and {Rezeau}}]{Sahraoui:2010b}
\bibinfo{author}{F.~{Sahraoui}}, \bibinfo{author}{M.~L. {Goldstein}},
  \bibinfo{author}{G.~{Belmont}}, \bibinfo{author}{P.~{Canu}},
  \bibinfo{author}{L.~{Rezeau}}, \bibinfo{title}{{Three Dimensional Anisotropic
  k Spectra of Turbulence at Subproton Scales in the Solar Wind}},
  \bibinfo{journal}{Phys.~Rev.~Lett.}
  \bibinfo{volume}{105}~(\bibinfo{number}{13}) (\bibinfo{year}{2010})
  \bibinfo{pages}{131101--+},
  \doi{\bibinfo{doi}{10.1103/PhysRevLett.105.131101}}.

\bibitem[{{Alexandrova} et~al.(2012){Alexandrova}, {Lacombe}, {Mangeney},
  {Grappin}, and {Maksimovic}}]{Alexandrova:2012}
\bibinfo{author}{O.~{Alexandrova}}, \bibinfo{author}{C.~{Lacombe}},
  \bibinfo{author}{A.~{Mangeney}}, \bibinfo{author}{R.~{Grappin}},
  \bibinfo{author}{M.~{Maksimovic}}, \bibinfo{title}{{Solar Wind Turbulent
  Spectrum at Plasma Kinetic Scales}}, \bibinfo{journal}{Astrophys.~J.}
  \bibinfo{volume}{760} \bibinfo{eid}{121},
  \doi{\bibinfo{doi}{10.1088/0004-637X/760/2/121}}.

\bibitem[{{Wicks} et~al.(2013){Wicks}, {Mallet}, {Horbury}, {Chen},
  {Schekochihin}, and {Mitchell}}]{Wicks:2013}
\bibinfo{author}{R.~T. {Wicks}}, \bibinfo{author}{A.~{Mallet}},
  \bibinfo{author}{T.~S. {Horbury}}, \bibinfo{author}{C.~H.~K. {Chen}},
  \bibinfo{author}{A.~A. {Schekochihin}}, \bibinfo{author}{J.~J. {Mitchell}},
  \bibinfo{title}{{Alignment and Scaling of Large-Scale Fluctuations in the
  Solar Wind}}, \bibinfo{journal}{Phys.~Rev.~Lett.}
  \bibinfo{volume}{110}~(\bibinfo{number}{2}) \bibinfo{eid}{025003},
  \doi{\bibinfo{doi}{10.1103/PhysRevLett.110.025003}}.

\bibitem[{{Horbury} et~al.(2008){Horbury}, Forman, and Oughton}]{Horbury:2008}
\bibinfo{author}{T.~S. {Horbury}}, \bibinfo{author}{M.~Forman},
  \bibinfo{author}{S.~Oughton}, \bibinfo{title}{Anisotropic Scaling of
  Magnetohydrodynamic Turbulence}, \bibinfo{journal}{Phys.~Rev.~Lett.}
  \bibinfo{volume}{101} (\bibinfo{year}{2008}) \bibinfo{pages}{175005},
  \doi{\bibinfo{doi}{10.1103/PhysRevLett.101.175005}}.

\bibitem[{{Podesta}(2009)}]{Podesta:2009a}
\bibinfo{author}{J.~J. {Podesta}}, \bibinfo{title}{{Dependence of Solar-Wind
  Power Spectra on the Direction of the Local Mean Magnetic Field}},
  \bibinfo{journal}{Astrophys.~J.} \bibinfo{volume}{698} (\bibinfo{year}{2009})
  \bibinfo{pages}{986--999}, \doi{\bibinfo{doi}{10.1088/0004-637X/698/2/986}}.

\bibitem[{{Wicks} et~al.(2010){Wicks}, {Horbury}, {Chen}, and
  {Schekochihin}}]{Wicks:2010a}
\bibinfo{author}{R.~T. {Wicks}}, \bibinfo{author}{T.~S. {Horbury}},
  \bibinfo{author}{C.~H.~K. {Chen}}, \bibinfo{author}{A.~A. {Schekochihin}},
  \bibinfo{title}{{Power and spectral index anisotropy of the entire inertial
  range of turbulence in the fast solar wind}},
  \bibinfo{journal}{Mon.~Not.~Roy.~Astron.~Soc.} \bibinfo{volume}{407}
  (\bibinfo{year}{2010}) \bibinfo{pages}{L31--L35},
  \doi{\bibinfo{doi}{10.1111/j.1745-3933.2010.00898.x}}.

\bibitem[{{Luo} and {Wu}(2010)}]{Luo:2010}
\bibinfo{author}{Q.~Y. {Luo}}, \bibinfo{author}{D.~J. {Wu}},
  \bibinfo{title}{{Observations of Anisotropic Scaling of Solar Wind
  Turbulence}}, \bibinfo{journal}{Astrophys.~J.~Lett.} \bibinfo{volume}{714}
  (\bibinfo{year}{2010}) \bibinfo{pages}{L138--L141},
  \doi{\bibinfo{doi}{10.1088/2041-8205/714/1/L138}}.

\bibitem[{{Chen} et~al.(2011){Chen}, {Mallet}, {Schekochihin}, {Horbury},
  {Wicks}, and {Bale}}]{Chen:2011}
\bibinfo{author}{C.~H.~K. {Chen}}, \bibinfo{author}{A.~{Mallet}},
  \bibinfo{author}{A.~A. {Schekochihin}}, \bibinfo{author}{T.~S. {Horbury}},
  \bibinfo{author}{R.~T. {Wicks}}, \bibinfo{author}{S.~D. {Bale}},
  \bibinfo{title}{{Three-Dimensional Structure of Solar Wind Turbulence}},
  \bibinfo{journal}{ArXiv e-prints} .

\bibitem[{{Narita} et~al.(2011){Narita}, {Gary}, {Saito}, {Glassmeier}, and
  {Motschmann}}]{Narita:2011}
\bibinfo{author}{Y.~{Narita}}, \bibinfo{author}{S.~P. {Gary}},
  \bibinfo{author}{S.~{Saito}}, \bibinfo{author}{K.-H. {Glassmeier}},
  \bibinfo{author}{U.~{Motschmann}}, \bibinfo{title}{{Dispersion relation
  analysis of solar wind turbulence}}, \bibinfo{journal}{Geophys.~Res.~Lett.}
  \bibinfo{volume}{38} (\bibinfo{year}{2011}) \bibinfo{pages}{L05101},
  \doi{\bibinfo{doi}{10.1029/2010GL046588}}.

\bibitem[{Cho and Vishniac(2000)}]{Cho:2000}
\bibinfo{author}{J.~Cho}, \bibinfo{author}{E.~T. Vishniac},
  \bibinfo{title}{{The Anisotropy of Magnetohydrodynamic Alfv\'enic
  Turbulence}}, \bibinfo{journal}{Astrophys.~J.} \bibinfo{volume}{539}
  (\bibinfo{year}{2000}) \bibinfo{pages}{273--282}.

\bibitem[{Maron and Goldreich(2001)}]{Maron:2001}
\bibinfo{author}{J.~Maron}, \bibinfo{author}{P.~Goldreich},
  \bibinfo{title}{Simulations of Incompressible Magnetohydrodynamic
  Turbulence}, \bibinfo{journal}{Astrophys.~J.} \bibinfo{volume}{554}
  (\bibinfo{year}{2001}) \bibinfo{pages}{1175--1196}.

\bibitem[{{Cho} and {Lazarian}(2002)}]{Cho:2002a}
\bibinfo{author}{J.~{Cho}}, \bibinfo{author}{A.~{Lazarian}},
  \bibinfo{title}{{Compressible Sub-Alfv{\'e}nic MHD Turbulence in Low-
  {$\beta$} Plasmas}}, \bibinfo{journal}{Phys.~Rev.~Lett.}
  \bibinfo{volume}{88}~(\bibinfo{number}{24}) \bibinfo{eid}{245001},
  \doi{\bibinfo{doi}{10.1103/PhysRevLett.88.245001}}.

\bibitem[{{Mason} et~al.(2006){Mason}, {Cattaneo}, and {Boldyrev}}]{Mason:2006}
\bibinfo{author}{J.~{Mason}}, \bibinfo{author}{F.~{Cattaneo}},
  \bibinfo{author}{S.~{Boldyrev}}, \bibinfo{title}{{Dynamic Alignment in Driven
  Magnetohydrodynamic Turbulence}}, \bibinfo{journal}{Phys.~Rev.~Lett.}
  \bibinfo{volume}{97}~(\bibinfo{number}{25}) (\bibinfo{year}{2006})
  \bibinfo{pages}{255002--+},
  \doi{\bibinfo{doi}{10.1103/PhysRevLett.97.255002}}.

\bibitem[{{Mason} et~al.(2011){Mason}, {Perez}, {Cattaneo}, and
  {Boldyrev}}]{Mason:2011}
\bibinfo{author}{J.~{Mason}}, \bibinfo{author}{J.~C. {Perez}},
  \bibinfo{author}{F.~{Cattaneo}}, \bibinfo{author}{S.~{Boldyrev}},
  \bibinfo{title}{{Extended Scaling Laws in Numerical Simulations of
  Magnetohydrodynamic Turbulence}}, \bibinfo{journal}{Astrophys.~J.~Lett.}
  \bibinfo{volume}{735} \bibinfo{eid}{L26},
  \doi{\bibinfo{doi}{10.1088/2041-8205/735/2/L26}}.

\bibitem[{{Perez} et~al.(2012){Perez}, {Mason}, {Boldyrev}, and
  {Cattaneo}}]{Perez:2012}
\bibinfo{author}{J.~C. {Perez}}, \bibinfo{author}{J.~{Mason}},
  \bibinfo{author}{S.~{Boldyrev}}, \bibinfo{author}{F.~{Cattaneo}},
  \bibinfo{title}{{On the Energy Spectrum of Strong Magnetohydrodynamic
  Turbulence}}, \bibinfo{journal}{Phys.~Rev.~X}
  \bibinfo{volume}{2}~(\bibinfo{number}{4}) \bibinfo{eid}{041005},
  \doi{\bibinfo{doi}{10.1103/PhysRevX.2.041005}}.

\bibitem[{{Howes} et~al.(2008{\natexlab{b}}){Howes}, {Dorland}, {Cowley},
  {Hammett}, {Quataert}, {Schekochihin}, and {Tatsuno}}]{Howes:2008a}
\bibinfo{author}{G.~G. {Howes}}, \bibinfo{author}{W.~{Dorland}},
  \bibinfo{author}{S.~C. {Cowley}}, \bibinfo{author}{G.~W. {Hammett}},
  \bibinfo{author}{E.~{Quataert}}, \bibinfo{author}{A.~A. {Schekochihin}},
  \bibinfo{author}{T.~{Tatsuno}}, \bibinfo{title}{{Kinetic Simulations of
  Magnetized Turbulence in Astrophysical Plasmas}},
  \bibinfo{journal}{Phys.~Rev.~Lett.}
  \bibinfo{volume}{100}~(\bibinfo{number}{6})
  (\bibinfo{year}{2008}{\natexlab{b}}) \bibinfo{pages}{065004},
  \doi{\bibinfo{doi}{10.1103/PhysRevLett.100.065004}}.

\bibitem[{{Salem} et~al.(2012){Salem}, {Howes}, {Sundkvist}, {Bale}, {Chaston},
  {Chen}, and {Mozer}}]{Salem:2012}
\bibinfo{author}{C.~S. {Salem}}, \bibinfo{author}{G.~G. {Howes}},
  \bibinfo{author}{D.~{Sundkvist}}, \bibinfo{author}{S.~D. {Bale}},
  \bibinfo{author}{C.~C. {Chaston}}, \bibinfo{author}{C.~H.~K. {Chen}},
  \bibinfo{author}{F.~S. {Mozer}}, \bibinfo{title}{{Identification of Kinetic
  Alfv{\'e}n Wave Turbulence in the Solar Wind}},
  \bibinfo{journal}{Astrophys.~J.~Lett.} \bibinfo{volume}{745}
  \bibinfo{eid}{L9}, \doi{\bibinfo{doi}{10.1088/2041-8205/745/1/L9}}.

\bibitem[{{TenBarge} et~al.(2012){TenBarge}, {Podesta}, {Klein}, and
  {Howes}}]{TenBarge:2012d}
\bibinfo{author}{J.~M. {TenBarge}}, \bibinfo{author}{J.~J. {Podesta}},
  \bibinfo{author}{K.~G. {Klein}}, \bibinfo{author}{G.~G. {Howes}},
  \bibinfo{title}{{Interpreting Magnetic Variance Anisotropy Measurements in
  the Solar Wind}}, \bibinfo{journal}{Astrophys.~J.} \bibinfo{volume}{753}
  \bibinfo{eid}{107}, \doi{\bibinfo{doi}{10.1088/0004-637X/753/2/107}}.

\bibitem[{{Howes} et~al.(2011{\natexlab{a}}){Howes}, {TenBarge}, {Dorland},
  {Quataert}, {Schekochihin}, {Numata}, and {Tatsuno}}]{Howes:2011b}
\bibinfo{author}{G.~G. {Howes}}, \bibinfo{author}{J.~M. {TenBarge}},
  \bibinfo{author}{W.~{Dorland}}, \bibinfo{author}{E.~{Quataert}},
  \bibinfo{author}{A.~A. {Schekochihin}}, \bibinfo{author}{R.~{Numata}},
  \bibinfo{author}{T.~{Tatsuno}}, \bibinfo{title}{{Gyrokinetic Simulations of
  Solar Wind Turbulence from Ion to Electron Scales}},
  \bibinfo{journal}{Phys.~Rev.~Lett.}
  \bibinfo{volume}{107}~(\bibinfo{number}{3})
  (\bibinfo{year}{2011}{\natexlab{a}}) \bibinfo{pages}{035004--+},
  \doi{\bibinfo{doi}{10.1103/PhysRevLett.107.035004}}.

\bibitem[{{TenBarge} et~al.(2013){TenBarge}, {Howes}, and
  {Dorland}}]{TenBarge:2012c}
\bibinfo{author}{J.~M. {TenBarge}}, \bibinfo{author}{G.~G. {Howes}},
  \bibinfo{author}{W.~{Dorland}}, \bibinfo{title}{{Collisionless Dissipation at
  Electron Scales in the Solar Wind}}, \bibinfo{journal}{Astrophys.~J.}
  \bibinfo{volume}{774} (\bibinfo{year}{2013}) \bibinfo{pages}{107},
  \bibinfo{note}{submitted}.

\bibitem[{{Cho} and {Lazarian}(2004)}]{Cho:2004}
\bibinfo{author}{J.~{Cho}}, \bibinfo{author}{A.~{Lazarian}},
  \bibinfo{title}{{The Anisotropy of Electron Magnetohydrodynamic Turbulence}},
  \bibinfo{journal}{Astrophys.~J.~Lett.} \bibinfo{volume}{615}
  (\bibinfo{year}{2004}) \bibinfo{pages}{L41--L44},
  \doi{\bibinfo{doi}{10.1086/425215}}.

\bibitem[{{Cho} and {Lazarian}(2009)}]{Cho:2009}
\bibinfo{author}{J.~{Cho}}, \bibinfo{author}{A.~{Lazarian}},
  \bibinfo{title}{{Simulations of Electron Magnetohydrodynamic Turbulence}},
  \bibinfo{journal}{Astrophys.~J.} \bibinfo{volume}{701} (\bibinfo{year}{2009})
  \bibinfo{pages}{236--252}, \doi{\bibinfo{doi}{10.1088/0004-637X/701/1/236}}.

\bibitem[{{TenBarge} and {Howes}(2012)}]{TenBarge:2012a}
\bibinfo{author}{J.~M. {TenBarge}}, \bibinfo{author}{G.~G. {Howes}},
  \bibinfo{title}{{Evidence of Critical Balance in Kinetic Alfven Wave
  Turbulence Simulations}}, \bibinfo{journal}{Phys.~Plasmas}
  \bibinfo{volume}{19}~(\bibinfo{number}{5}) (\bibinfo{year}{2012})
  \bibinfo{pages}{055901}.

\bibitem[{{Perez} and {Boldyrev}(2008)}]{Perez:2008}
\bibinfo{author}{J.~C. {Perez}}, \bibinfo{author}{S.~{Boldyrev}},
  \bibinfo{title}{{On Weak and Strong Magnetohydrodynamic Turbulence}},
  \bibinfo{journal}{Astrophys.~J.~Lett.} \bibinfo{volume}{672}
  (\bibinfo{year}{2008}) \bibinfo{pages}{L61--L64},
  \doi{\bibinfo{doi}{10.1086/526342}}.

\bibitem[{{Perez} and {Boldyrev}(2010)}]{Perez:2010a}
\bibinfo{author}{J.~C. {Perez}}, \bibinfo{author}{S.~{Boldyrev}},
  \bibinfo{title}{{Strong magnetohydrodynamic turbulence with cross helicity}},
  \bibinfo{journal}{Physics of Plasmas}
  \bibinfo{volume}{17}~(\bibinfo{number}{5}) (\bibinfo{year}{2010})
  \bibinfo{pages}{055903}, \doi{\bibinfo{doi}{10.1063/1.3396370}}.

\bibitem[{{Gary} et~al.(2010){Gary}, {Saito}, and {Narita}}]{Gary:2010}
\bibinfo{author}{S.~P. {Gary}}, \bibinfo{author}{S.~{Saito}},
  \bibinfo{author}{Y.~{Narita}}, \bibinfo{title}{{Whistler Turbulence
  Wavevector Anisotropies: Particle-in-cell Simulations}},
  \bibinfo{journal}{Astrophys.~J.} \bibinfo{volume}{716} (\bibinfo{year}{2010})
  \bibinfo{pages}{1332--1335},
  \doi{\bibinfo{doi}{10.1088/0004-637X/716/2/1332}}.

\bibitem[{{Markovskii} and {Vasquez}(2011)}]{Markovskii:2011}
\bibinfo{author}{S.~A. {Markovskii}}, \bibinfo{author}{B.~J. {Vasquez}},
  \bibinfo{title}{{A Short-timescale Channel of Dissipation of the Strong Solar
  Wind Turbulence}}, \bibinfo{journal}{Astrophys.~J.} \bibinfo{volume}{739}
  \bibinfo{eid}{22}, \doi{\bibinfo{doi}{10.1088/0004-637X/739/1/22}}.

\bibitem[{{Vasquez} and {Markovskii}(2012)}]{Vasquez:2012}
\bibinfo{author}{B.~J. {Vasquez}}, \bibinfo{author}{S.~A. {Markovskii}},
  \bibinfo{title}{{Velocity Power Spectra from Cross-field Turbulence in the
  Proton Kinetic Regime}}, \bibinfo{journal}{Astrophys.~J.}
  \bibinfo{volume}{747} \bibinfo{eid}{19},
  \doi{\bibinfo{doi}{10.1088/0004-637X/747/1/19}}.

\bibitem[{{Donato} et~al.(2012){Donato}, {Servidio}, {Dmitruk}, {Carbone},
  {Shay}, {Cassak}, and {Matthaeus}}]{Donato:2012}
\bibinfo{author}{S.~{Donato}}, \bibinfo{author}{S.~{Servidio}},
  \bibinfo{author}{P.~{Dmitruk}}, \bibinfo{author}{V.~{Carbone}},
  \bibinfo{author}{M.~A. {Shay}}, \bibinfo{author}{P.~A. {Cassak}},
  \bibinfo{author}{W.~H. {Matthaeus}}, \bibinfo{title}{{Reconnection events in
  two-dimensional Hall magnetohydrodynamic turbulence}},
  \bibinfo{journal}{Phys.~Plasmas} \bibinfo{volume}{19}~(\bibinfo{number}{9})
  (\bibinfo{year}{2012}) \bibinfo{pages}{092307},
  \doi{\bibinfo{doi}{10.1063/1.4754151}}.

\bibitem[{{Wan} et~al.(2012){Wan}, {Matthaeus}, {Karimabadi}, {Roytershteyn},
  {Shay}, {Wu}, {Daughton}, {Loring}, and {Chapman}}]{Wan:2012a}
\bibinfo{author}{M.~{Wan}}, \bibinfo{author}{W.~H. {Matthaeus}},
  \bibinfo{author}{H.~{Karimabadi}}, \bibinfo{author}{V.~{Roytershteyn}},
  \bibinfo{author}{M.~{Shay}}, \bibinfo{author}{P.~{Wu}},
  \bibinfo{author}{W.~{Daughton}}, \bibinfo{author}{B.~{Loring}},
  \bibinfo{author}{S.~C. {Chapman}}, \bibinfo{title}{{Intermittent Dissipation
  at Kinetic Scales in Collisionless Plasma Turbulence}},
  \bibinfo{journal}{Phys.~Rev.~Lett.}
  \bibinfo{volume}{109}~(\bibinfo{number}{19}) \bibinfo{eid}{195001},
  \doi{\bibinfo{doi}{10.1103/PhysRevLett.109.195001}}.

\bibitem[{{Saito} et~al.(2008){Saito}, {Gary}, {Li}, and {Narita}}]{Saito:2008}
\bibinfo{author}{S.~{Saito}}, \bibinfo{author}{S.~P. {Gary}},
  \bibinfo{author}{H.~{Li}}, \bibinfo{author}{Y.~{Narita}},
  \bibinfo{title}{{Whistler turbulence: Particle-in-cell simulations}},
  \bibinfo{journal}{Phys.~Plasmas} \bibinfo{volume}{15}~(\bibinfo{number}{10})
  (\bibinfo{year}{2008}) \bibinfo{pages}{102305--+},
  \doi{\bibinfo{doi}{10.1063/1.2997339}}.

\bibitem[{{Numata} et~al.(2010){Numata}, {Howes}, {Tatsuno}, {Barnes}, and
  {Dorland}}]{Numata:2010}
\bibinfo{author}{R.~{Numata}}, \bibinfo{author}{G.~G. {Howes}},
  \bibinfo{author}{T.~{Tatsuno}}, \bibinfo{author}{M.~{Barnes}},
  \bibinfo{author}{W.~{Dorland}}, \bibinfo{title}{{AstroGK: Astrophysical
  Gyrokinetics Code}}, \bibinfo{journal}{J.~Comp.~Phys.} \bibinfo{volume}{229}
  (\bibinfo{year}{2010}) \bibinfo{pages}{9347--9372},
  \doi{\bibinfo{doi}{10.1016/j.jcp.2010.09.006}}.

\bibitem[{{Saito} et~al.(2010){Saito}, {Gary}, and {Narita}}]{Saito:2010}
\bibinfo{author}{S.~{Saito}}, \bibinfo{author}{S.~P. {Gary}},
  \bibinfo{author}{Y.~{Narita}}, \bibinfo{title}{{Wavenumber spectrum of
  whistler turbulence: Particle-in-cell simulation}},
  \bibinfo{journal}{Phys.~Plasmas} \bibinfo{volume}{17}~(\bibinfo{number}{12})
  (\bibinfo{year}{2010}) \bibinfo{pages}{122316--+},
  \doi{\bibinfo{doi}{10.1063/1.3526602}}.

\bibitem[{{Parashar} et~al.(2010){Parashar}, {Servidio}, {Breech}, {Shay}, and
  {Matthaeus}}]{Parashar:2010}
\bibinfo{author}{T.~N. {Parashar}}, \bibinfo{author}{S.~{Servidio}},
  \bibinfo{author}{B.~{Breech}}, \bibinfo{author}{M.~A. {Shay}},
  \bibinfo{author}{W.~H. {Matthaeus}}, \bibinfo{title}{{Kinetic driven
  turbulence: Structure in space and time}}, \bibinfo{journal}{Phys.~Plasmas}
  \bibinfo{volume}{17}~(\bibinfo{number}{10}) (\bibinfo{year}{2010})
  \bibinfo{pages}{102304}, \doi{\bibinfo{doi}{10.1063/1.3486537}}.

\bibitem[{{Chang} et~al.(2011){Chang}, {Peter Gary}, and {Wang}}]{Chang:2011}
\bibinfo{author}{O.~{Chang}}, \bibinfo{author}{S.~{Peter Gary}},
  \bibinfo{author}{J.~{Wang}}, \bibinfo{title}{{Whistler turbulence forward
  cascade: Three-dimensional particle-in-cell simulations}},
  \bibinfo{journal}{Geophys.~Res.~Lett.} \bibinfo{volume}{38}
  \bibinfo{eid}{L22102}, \doi{\bibinfo{doi}{10.1029/2011GL049827}}.

\bibitem[{{Gary} et~al.(2012){Gary}, {Chang}, and {Wang}}]{Gary:2012}
\bibinfo{author}{S.~P. {Gary}}, \bibinfo{author}{O.~{Chang}},
  \bibinfo{author}{J.~{Wang}}, \bibinfo{title}{{Forward Cascade of Whistler
  Turbulence: Three-dimensional Particle-in-cell Simulations}},
  \bibinfo{journal}{Astrophys.~J.} \bibinfo{volume}{755} \bibinfo{eid}{142},
  \doi{\bibinfo{doi}{10.1088/0004-637X/755/2/142}}.

\bibitem[{Higdon(1984)}]{Higdon:1984a}
\bibinfo{author}{J.~C. Higdon}, \bibinfo{title}{Density Fluctuations in the
  Interstellar Medium: Evidence for Anisotropic Magnetogasdynamic Turbulence I.
  Model and Astrophysical Sites}, \bibinfo{journal}{Astrophys.~J.}
  \bibinfo{volume}{285} (\bibinfo{year}{1984}) \bibinfo{pages}{109--123}.

\bibitem[{{Beresnyak} and {Lazarian}(2009)}]{Beresnyak:2009}
\bibinfo{author}{A.~{Beresnyak}}, \bibinfo{author}{A.~{Lazarian}},
  \bibinfo{title}{{Comparison of Spectral Slopes of Magnetohydrodynamic and
  Hydrodynamic Turbulence and Measurements of Alignment Effects}},
  \bibinfo{journal}{Astrophys.~J.} \bibinfo{volume}{702} (\bibinfo{year}{2009})
  \bibinfo{pages}{1190--1198},
  \doi{\bibinfo{doi}{10.1088/0004-637X/702/2/1190}}.

\bibitem[{{Elsasser}(1950)}]{Elsasser:1950}
\bibinfo{author}{W.~M. {Elsasser}}, \bibinfo{title}{{The Hydromagnetic
  Equations}}, \bibinfo{journal}{Physical Review} \bibinfo{volume}{79}
  (\bibinfo{year}{1950}) \bibinfo{pages}{183--183},
  \doi{\bibinfo{doi}{10.1103/PhysRev.79.183}}.

\bibitem[{{Parashar} et~al.(2011){Parashar}, {Servidio}, {Shay}, {Breech}, and
  {Matthaeus}}]{Parashar:2011}
\bibinfo{author}{T.~N. {Parashar}}, \bibinfo{author}{S.~{Servidio}},
  \bibinfo{author}{M.~A. {Shay}}, \bibinfo{author}{B.~{Breech}},
  \bibinfo{author}{W.~H. {Matthaeus}}, \bibinfo{title}{{Effect of driving
  frequency on excitation of turbulence in a kinetic plasma}},
  \bibinfo{journal}{Phys.~Plasmas} \bibinfo{volume}{18}~(\bibinfo{number}{9})
  (\bibinfo{year}{2011}) \bibinfo{pages}{092302--+},
  \doi{\bibinfo{doi}{10.1063/1.3630926}}.

\bibitem[{{Lemons}(2002)}]{Lemons:2002}
\bibinfo{author}{D.~S. {Lemons}}, \bibinfo{title}{{An Introdction to Stochastic
  Processes in Physics}}, \bibinfo{publisher}{Baltimore: Johns Hopkins
  University Press}, \bibinfo{year}{2002}.

\bibitem[{{Langevin}(1908)}]{Langevin:1908}
\bibinfo{author}{P.~{Langevin}}, \bibinfo{title}{{On the Theory of Brownian
  Motion}}, \bibinfo{journal}{C. R. Acad. Sci. (Paris)} \bibinfo{volume}{146}
  (\bibinfo{year}{1908}) \bibinfo{pages}{530--533}.

\bibitem[{{Frieman} and {Chen}(1982)}]{Frieman:1982}
\bibinfo{author}{E.~A. {Frieman}}, \bibinfo{author}{L.~{Chen}},
  \bibinfo{title}{{Nonlinear gyrokinetic equations for low-frequency
  electromagnetic waves in general plasma equilibria}},
  \bibinfo{journal}{Phys.~Fluids} \bibinfo{volume}{25} (\bibinfo{year}{1982})
  \bibinfo{pages}{502--508}.

\bibitem[{{Howes} et~al.(2006){Howes}, {Cowley}, {Dorland}, {Hammett},
  {Quataert}, and {Schekochihin}}]{Howes:2006}
\bibinfo{author}{G.~G. {Howes}}, \bibinfo{author}{S.~C. {Cowley}},
  \bibinfo{author}{W.~{Dorland}}, \bibinfo{author}{G.~W. {Hammett}},
  \bibinfo{author}{E.~{Quataert}}, \bibinfo{author}{A.~A. {Schekochihin}},
  \bibinfo{title}{{Astrophysical Gyrokinetics: Basic Equations and Linear
  Theory}}, \bibinfo{journal}{Astrophys.~J.} \bibinfo{volume}{651}
  (\bibinfo{year}{2006}) \bibinfo{pages}{590--614},
  \doi{\bibinfo{doi}{10.1086/506172}}.

\bibitem[{{Abel} et~al.(2008){Abel}, {Barnes}, {Cowley}, {Dorland}, and
  {Schekochihin}}]{Abel:2008}
\bibinfo{author}{I.~G. {Abel}}, \bibinfo{author}{M.~{Barnes}},
  \bibinfo{author}{S.~C. {Cowley}}, \bibinfo{author}{W.~{Dorland}},
  \bibinfo{author}{A.~A. {Schekochihin}}, \bibinfo{title}{{Linearized model
  Fokker-Planck collision operators for gyrokinetic simulations. I. Theory}},
  \bibinfo{journal}{Phys.~Plasmas} \bibinfo{volume}{15}~(\bibinfo{number}{12})
  (\bibinfo{year}{2008}) \bibinfo{pages}{122509--+},
  \doi{\bibinfo{doi}{10.1063/1.3046067}}.

\bibitem[{{Barnes} et~al.(2009){Barnes}, {Abel}, {Dorland}, {Ernst}, {Hammett},
  {Ricci}, {Rogers}, {Schekochihin}, and {Tatsuno}}]{Barnes:2009}
\bibinfo{author}{M.~{Barnes}}, \bibinfo{author}{I.~G. {Abel}},
  \bibinfo{author}{W.~{Dorland}}, \bibinfo{author}{D.~R. {Ernst}},
  \bibinfo{author}{G.~W. {Hammett}}, \bibinfo{author}{P.~{Ricci}},
  \bibinfo{author}{B.~N. {Rogers}}, \bibinfo{author}{A.~A. {Schekochihin}},
  \bibinfo{author}{T.~{Tatsuno}}, \bibinfo{title}{{Linearized model
  Fokker-Planck collision operators for gyrokinetic simulations. II. Numerical
  implementation and tests}}, \bibinfo{journal}{Phys.~Plasmas}
  \bibinfo{volume}{16}~(\bibinfo{number}{7}) (\bibinfo{year}{2009})
  \bibinfo{pages}{072107--+}, \doi{\bibinfo{doi}{10.1063/1.3155085}}.

\bibitem[{{Howes} et~al.(2011{\natexlab{b}}){Howes}, {TenBarge}, and
  {Dorland}}]{Howes:2011c}
\bibinfo{author}{G.~G. {Howes}}, \bibinfo{author}{J.~M. {TenBarge}},
  \bibinfo{author}{W.~{Dorland}}, \bibinfo{title}{{A weakened cascade model for
  turbulence in astrophysical plasmas}}, \bibinfo{journal}{Phys.~Plasmas}
  \bibinfo{volume}{18}~(\bibinfo{number}{10})
  (\bibinfo{year}{2011}{\natexlab{b}}) \bibinfo{pages}{102305},
  \doi{\bibinfo{doi}{10.1063/1.3646400}}.

\bibitem[{Montgomery and Turner(1981)}]{Montgomery:1981}
\bibinfo{author}{D.~Montgomery}, \bibinfo{author}{L.~Turner},
  \bibinfo{title}{Anisotropic Magnetohydrodynamic Turbulence in a Strong
  External Magnetic Field}, \bibinfo{journal}{Phys.~Fluids}
  \bibinfo{volume}{24} (\bibinfo{year}{1981}) \bibinfo{pages}{825--831}.

\bibitem[{{Galtier} and {Bhattacharjee}(2003)}]{Galtier:2003}
\bibinfo{author}{S.~{Galtier}}, \bibinfo{author}{A.~{Bhattacharjee}},
  \bibinfo{title}{{Anisotropic weak whistler wave turbulence in electron
  magnetohydrodynamics}}, \bibinfo{journal}{Phys.~Plasmas} \bibinfo{volume}{10}
  (\bibinfo{year}{2003}) \bibinfo{pages}{3065--3076},
  \doi{\bibinfo{doi}{10.1063/1.1584433}}.

\bibitem[{Biskmap et~al.(1999)Biskmap, Schwarz, Zeiler, Celani, and
  Drake}]{Biskamp:1999}
\bibinfo{author}{D.~Biskmap}, \bibinfo{author}{E.~Schwarz},
  \bibinfo{author}{A.~Zeiler}, \bibinfo{author}{A.~Celani},
  \bibinfo{author}{J.~F. Drake}, \bibinfo{title}{Electron Magnetohydrodynamic
  Turbulence}, \bibinfo{journal}{Phys.~Plasmas} \bibinfo{volume}{6}
  (\bibinfo{year}{1999}) \bibinfo{pages}{751--758}.

\bibitem[{{Dastgeer} et~al.(2000){Dastgeer}, {Das}, {Kaw}, and
  {Diamond}}]{Dastgeer:2000}
\bibinfo{author}{S.~{Dastgeer}}, \bibinfo{author}{A.~{Das}},
  \bibinfo{author}{P.~{Kaw}}, \bibinfo{author}{P.~H. {Diamond}},
  \bibinfo{title}{{Whistlerization and anisotropy in two-dimensional electron
  magnetohydrodynamic turbulence}}, \bibinfo{journal}{Phys.~Plasmas}
  \bibinfo{volume}{7} (\bibinfo{year}{2000}) \bibinfo{pages}{571--579},
  \doi{\bibinfo{doi}{10.1063/1.873843}}.

\bibitem[{{Svidzinski} et~al.(2009){Svidzinski}, {Li}, {Rose}, {Albright}, and
  {Bowers}}]{Svidzinski:2009}
\bibinfo{author}{V.~A. {Svidzinski}}, \bibinfo{author}{H.~{Li}},
  \bibinfo{author}{H.~A. {Rose}}, \bibinfo{author}{B.~J. {Albright}},
  \bibinfo{author}{K.~J. {Bowers}}, \bibinfo{title}{{Particle in cell
  simulations of fast magnetosonic wave turbulence in the ion cyclotron
  frequency range}}, \bibinfo{journal}{Phys.~Plasmas}
  \bibinfo{volume}{16}~(\bibinfo{number}{12}) (\bibinfo{year}{2009})
  \bibinfo{pages}{122310--+}, \doi{\bibinfo{doi}{10.1063/1.3274559}}.

\bibitem[{{Tatsuno} et~al.(2010){Tatsuno}, {Barnes}, {Cowley}, {Dorland},
  {Howes}, {Numata}, {Plunk}, and {Schekochihin}}]{Tatsuno:2010}
\bibinfo{author}{T.~{Tatsuno}}, \bibinfo{author}{M.~{Barnes}},
  \bibinfo{author}{S.~C. {Cowley}}, \bibinfo{author}{W.~{Dorland}},
  \bibinfo{author}{G.~G. {Howes}}, \bibinfo{author}{R.~{Numata}},
  \bibinfo{author}{G.~G. {Plunk}}, \bibinfo{author}{A.~A. {Schekochihin}},
  \bibinfo{title}{{Gyrokinetic simulation of entropy cascade in two-dimensional
  electrostatic turbulence}}, \bibinfo{journal}{J. Plasma Fusion Res.}
  \bibinfo{note}{Accepted}.

\bibitem[{{Hunana} et~al.(2011){Hunana}, {Laveder}, {Passot}, {Sulem}, and
  {Borgogno}}]{Hunana:2011}
\bibinfo{author}{P.~{Hunana}}, \bibinfo{author}{D.~{Laveder}},
  \bibinfo{author}{T.~{Passot}}, \bibinfo{author}{P.~L. {Sulem}},
  \bibinfo{author}{D.~{Borgogno}}, \bibinfo{title}{{Reduction of
  Compressibility and Parallel Transfer by Landau Damping in Turbulent
  Magnetized Plasmas}}, \bibinfo{journal}{Astrophys.~J.} \bibinfo{volume}{743}
  \bibinfo{eid}{128}, \doi{\bibinfo{doi}{10.1088/0004-637X/743/2/128}}.

\bibitem[{{Haugen} et~al.(2003){Haugen}, {Brandenburg}, and
  {Dobler}}]{Haugen:2003}
\bibinfo{author}{N.~E.~L. {Haugen}}, \bibinfo{author}{A.~{Brandenburg}},
  \bibinfo{author}{W.~{Dobler}}, \bibinfo{title}{{Is Nonhelical Hydromagnetic
  Turbulence Peaked at Small Scales?}}, \bibinfo{journal}{Astrophys.~J.~Lett.}
  \bibinfo{volume}{597} (\bibinfo{year}{2003}) \bibinfo{pages}{L141--L144},
  \doi{\bibinfo{doi}{10.1086/380189}}.

\bibitem[{{M{\"u}ller} and {Grappin}(2005)}]{Muller:2005}
\bibinfo{author}{W.-C. {M{\"u}ller}}, \bibinfo{author}{R.~{Grappin}},
  \bibinfo{title}{{Spectral Energy Dynamics in Magnetohydrodynamic
  Turbulence}}, \bibinfo{journal}{Phys.~Rev.~Lett.}
  \bibinfo{volume}{95}~(\bibinfo{number}{11}) \bibinfo{eid}{114502},
  \doi{\bibinfo{doi}{10.1103/PhysRevLett.95.114502}}.

\end{thebibliography}

\end{document}